\documentclass{aa} 
\usepackage{longtable}
\usepackage{latexsym}
\usepackage{amssymb}
\usepackage{lscape}
\usepackage[]{natbib}

\usepackage{graphicx}
\usepackage{txfonts}
\title{H$\alpha3$: an H$\alpha$ imaging survey of HI selected galaxies from ALFALFA\thanks{Based on observations 
taken at the  observatory of San Pedro Martir (Baja California, Mexico), belonging to the
Mexican Observatorio Astron\'omico Nacional.}}

\subtitle{III.  Nurture builds up the Hubble sequence in the Great Wall}

\author{G. Gavazzi \inst{1}
\and G. Savorgnan \inst{1,2}
\and M. Fossati \inst{3,1}
\and M. Dotti \inst{1}
\and M. Fumagalli \inst{4,5}\thanks {Hubble Fellow}
\and A. Boselli \inst{6}
\and L. Guti\'errez \inst{7}
\and H. Hern\'andez Toledo \inst{8}
\and R. Giovanelli \inst{9}
\and M.P. Haynes \inst{9}
}
\authorrunning{G. Gavazzi et al.}
\titlerunning{H$\alpha3$: H$\alpha$ imaging survey of HI selected galaxies from ALFALFA}
\institute{Universit\`a degli Studi di Milano-Bicocca, Piazza della Scienza 3, 20126 Milano, Italy\\
\email {giuseppe.gavazzi@mib.infn.it}
\and
Centre for Astrophysics and Supercomputing, Swinburne University of Technology, Hawthorn, Victoria 3122, Australia\\
\email gsavorgn@astro.swin.edu.au
\and
Max-Planck-Institut f{\"u}r Extraterrestrische Physik, Giessenbachstrasse, D-85748 Garching, Germany\\
\email {mfossati@mpe.mpg.de}
\and
Carnegie Observatories, 813 Santa Barbara Street, Pasadena, CA 91101, USA\\
\email {mfumagalli@obs.carnegiescience.edu}
\and
Department of Astrophysics, Princeton University, Princeton, NJ 08544-1001, USA
\and
Aix Marseille Universit\'e, CNRS, LAM (Laboratoire d'Astrophysique de Marseille) UMR 7326, 13388, Marseille, France\\
\email {alessandro.boselli@oamp.fr}
\and
Instituto de Astronom\'ia, Universidad Nacional Aut\'onoma de M\'exico, 
Carretera Tijuana-Ensenada, km 103, 22860 Ensenada, B.C., M\'exico.\\
\email {leonel@astro.unam.mx}
\and
Instituto de Astronom\'ia, Universidad Nacional Aut\'onoma de M\'exico, 
Apartado Postal 70-264, 04510 M\'exico D.F., M\'exico.\\
\email {hector@astroscu.unam.mx}
\and
Center for Radiophysics and Space Research, Space Science Building, Ithaca, NY, 14853\\
\email {haynes@astro.cornell.edu, riccardo@astro.cornell.edu}
}

\begin{document}
\date{received 21/12/2012; accepted 10/3/2013}

 
  \abstract
         {We present the analysis of H$\alpha3$, 
           an H$\alpha$ narrow-band imaging follow-up survey of 
           galaxies selected from the HI Arecibo Legacy Fast ALFA Survey (ALFALFA) 
           in the Coma supercluster.
       }
	{Taking advantage of H$\alpha3$, which provides the complete census 
         of the recent star formation in HI-rich galaxies in the local universe,
	 we explored the hypothesis that a morphological sequence of galaxies 
	 of progressively earlier type and lower gas-content  exists in the neighborhood of a rich cluster of galaxies 
	 such as Coma, with a
	 specific star formation activity that decreases with increasing local galaxy density and velocity dispersion.
        }
         {By using the H$\alpha$ hydrogen recombination line as a tracer of the
           "instantaneous" star formation, complemented with optical colors from SDSS, we  
           investigated the relationships between atomic neutral gas and newly formed stars
           in different local galaxy density intervals, for many morphological types, 
	   and over a wide range of stellar masses ($10^{9}$ to $10^{11.5}$ M$_\odot$).}
  {In the dwarf regime ($8.5<\log(M_{*}/$M$_{\odot}) < 9.5$) we identify a four-step
  sequence of galaxies with progressively redder  colors (corrected for dust extinction), i.e., of
  decreasing specific star formation, from (1) HI-rich late-type galaxies (LTGs) belonging to the blue cloud
  that exhibit  extended plus nuclear star formation, (2) $\sim 0.1$ mag redder, HI-poor
  LTGs with nuclear star formation only, (3) $\sim 0.35$ mag redder, HI-poor
  galaxies without  either extended or nuclear star formation, but with nuclear
  post-star-burst (PSB) signature, (4) $\sim 0.5$ mag redder early-type galaxies (ETGs)  that belong to the red sequence,  
  and show no gas or star formation on all scales.  Along this sequence the quenching of
  the star formation proceeds radially outside-in.  
  The progression toward redder colors found along this "morphological" (gas content) sequence 
  is comparable to the one obtained from increasing the local galaxy density, from
  cosmic filaments (1 2), to the rich clusters (2 3 4).} 
 {In the dwarf regime we find evidence for an evolution of HI-rich LTGs into ETGs through
  HI-poor LTGs and PSB galaxies driven by the environment.  We identify ram-pressure as the 
  mechanism most likely responsible for this transformation. 
  We conclude that infall of galaxies has proceeded for the last 7.5 Gyr,
  building up the Coma cluster at a rate of approximately 100 galaxies with
  $\log(M_{*}/$M$_{\odot}) > 9.0$ per Gyr.\\
  }
         

   \maketitle

%

\section{Introduction}

Since the advent of the Sloan Digital Sky Survey (SDSS, York et al. 2000), which revolutionized research in astronomy
at the turn of the millennium, surveys carried out at frequencies other than optical were designed 
with an  extent, depth, and photometric and astrometric qualities comparable with  SDSS.\\
One such ambitious survey is ALFALFA (Giovanelli et al.  2005), a blind HI sky survey that just ended at Arecibo (October 2012), 
aimed at obtaining the census of HI sources within  7000 sq degrees of the sky accessible from Arecibo, with 
an rms noise near 2 mJy after Hanning smoothing to 10 $\rm km ~s^{-1}$
(see Haynes et al. 2011), corresponding to $10^{7.7}~\rm M_\odot$ of HI at the distance of Virgo and $10^{9}~\rm M_\odot$ at the distance of Coma.
A catalog listing 40\% of the whole ALFALFA sources available so far ($\alpha.40$ catalog) has been published by Haynes et al. (2011). 

The less extensive H$\alpha3$ 
is an  H$\alpha$ narrow-band imaging survey, ongoing at the 2.1m telescope of				   
the San Pedro Martir (SPM) Observatory, aimed at following-up all nearby galaxies that are selected from ALFALFA in 
the redshift  range ($-1000<cz<3000 ~\rm km~s^{-1}$) of the Local (Gavazzi et al. 2012) and Coma superclusters ($3900<cz<9500 ~\rm km~s^{-1}$).
The sensitivity of H$\alpha3$ is such that galaxies with star formation  rate (SFR)    
in excess of $\sim0.1~\rm M_\odot~yr^{-1}$ are sampled up to the distance of Coma (100 Mpc).					   
The three mentioned surveys provide us with the necessary ingredients for studying one of the most obvious, yet
still unsettled questions of galaxy evolution: the transformation of primeval atomic gas (HI) into $\rm H_2$, its
fragmentation in molecular clouds, and the birth of new stars that contribute to the formation and evolution of galaxies.
So far, the census of the SFR in the local Universe traced by hydrogen recombination lines					   
has been determined on HI selected surveys, such as SINGG selected from HIPASS  						   
(Meurer et al. 2006), or optically selected ones										   
(Boselli et al. 2001, Gavazzi et al. 2002b, 2006, James et al. 2004,								   
Kennicutt et al. 2008, Lee et al. 2007, 2009, Bothwell et al. 2009), or in optically selected galaxies 
with stellar mass larger than $10^{10}~\rm M_\odot$
(e.g. GASS, the GALEX Arecibo SDSS Survey, Catinella et al. 2010, Schiminovich et al. 2010, Fabello et al. 2012).										   
The combined sensitivities of SDSS, ALFALFA, and H$\alpha3$ are such that the present analysis can be extended
to dwarf galaxies with stellar masses as small as $10^{8}~\rm M_\odot$. 
Combining the $\alpha$.40 catalog (Haynes et al. 2011) with the catalog provided by the legacy program carried 
out with the Galaxy Evolution Explorer (GALEX), 
Huang et al. (2012) carried out the most recent investigation of the scaling relations between HI gas, 
star formation\footnote{We emphasize
that the SFR in normal galaxies can be estimated from the UV provided that
this remained constant over a time scale $t<10^8$ yr.
In contrast if the SFR   changed during a time scale $10^7<t<10^8$ yr, 
which is most likely the case of cluster and dwarf galaxies,
the most appropriate SFR indicator is H$\alpha$ (Boselli et al. 2009).}, and stellar mass.
They did not touch on the debated question whether the transformation of gas into stars which has provided the build-up 
of the Hubble sequence over the cosmic time,
depends on the environment (nurture) or is a purely ``genetic''  process (nature), which is the focus of the present work.
We follow the line traced by Boselli et al. (2008) and by Paper II (Gavazzi et al. 2013a), 
who found evidence that a strong role is played by the 
environment in the Virgo cluster for the transformation of star-forming dwarf galaxies of late type into quiescent dwarf elliptical galaxies
due to the fast ablation of the interstellar gas caused by the ram-pressure mechanism (Gunn \& Gott 1972), with consequent  
suppression of the star formation. 
Moreover, we continue on this line, following Gavazzi et al. (2010), who found
support for these ideas by studying the population of late-type galaxies surrounding the Coma cluster using photometry
and spectroscopy from the SDSS to separate the stellar populations in galaxies as faint as $M_i$=-17.5 mag.
The Coma supercluster consists of a main filament of galaxies running perpendicularly to the line of sight
across several hundreds of Mpc, with a distribution of galaxies spanning a significant density contrast. Its very shape, 
along with the new available data on the HI gas content from ALFALFA and the measurements of the star formation provided by  H$\alpha3$,
offer a unique opportunity of pursuing this question in the present Paper III of the H$\alpha3$ series.

Paper III is structured as follows.
The optical- and radio-selected samples used in the analysis are defined and discussed in Section 2.
Section 3 contains the evidence of the environmental dependency of the gas content and of the star formation activity
(both in the disk and circumnuclear). Discussion and summary are given in Section 4. 
A method for correcting the optical colors for internal extinction is obtained in Appendix A.
The scaling relations between HI mass, stellar mass and the star formation in Coma compared to the Local Supercluster are discussed
in Appendix B.

The H$\alpha3$ observations of the Local Supercluster have been the subject of Paper I of this series (Gavazzi et al. 2012).
Paper II (Gavazzi et al. 2013a) outlined the analysis, including the study of the scaling relations between HI mass, stellar mass, star formation,
and the environmental conditions.
Paper IV (Fossati et al. 2013) will analyze  the structural properties of galaxies in the Local and Coma superclusters, 
and Paper V (Gavazzi et al. 2013b) will contain the H$\alpha3$ data (fluxes and images) in the Coma supercluster.

\begin{figure*}[!t]
\centering
\includegraphics[width=19cm, trim = 0cm 1cm 0cm 0cm]{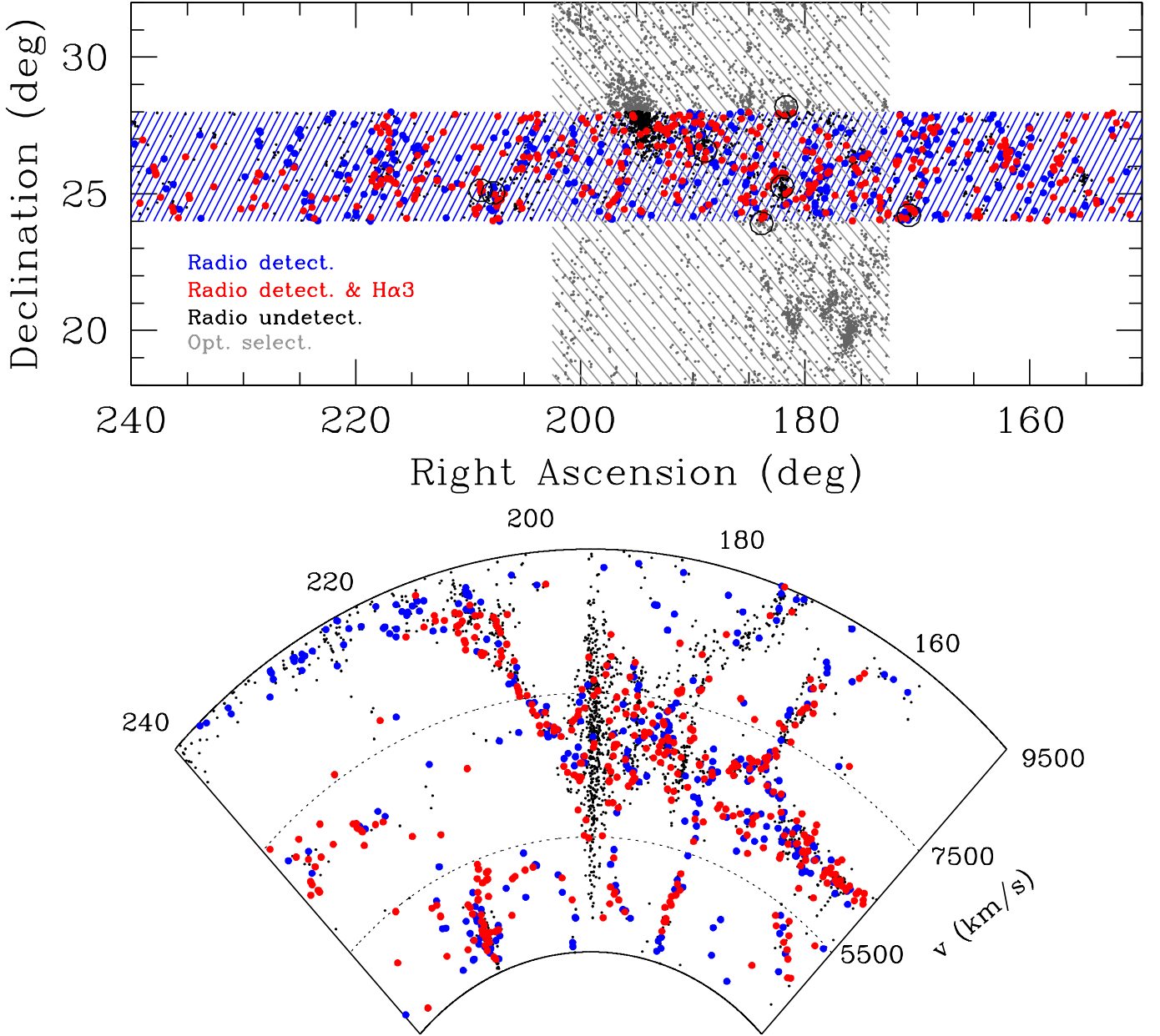}
 \caption{
 Top panel: sky distribution (the R.A. scale is compressed) of the Coma supercluster area studied in this work,
 constituted by the union of the traditional area occupied by the Coma supercluster  $\rm 172^o < R.A. <202^o$;
 $18^o< Dec. <32^o$ (region 1, gray shaded) with
 the elongated region $\rm 150^o < R.A. <240^o\, ; \,24^o< Dec. <28^o$ (region 2, blue shaded) covered by ALFALFA (Haynes et al. 2011).
 In both regions only galaxies in the interval $3900<cz<9500$  $\rm ~km~s^{-1}$ are considered. 
 Sample 1  includes 5026 galaxies of all morphological types optically selected from SDSS (small black dots) 
 in the union of regions 1+2.  
 Sample 2 includes 683 galaxies HI selected from ALFALFA (blue symbols).
 Sample 3  (a subset of sample 2) is composed of 
 383 galaxies followed-up with  H$\alpha$ observations as part of H$\alpha3$ (red symbols).
 Region 1 will be fully covered by ALFALFA in the near future. Currently, only the brightest galaxies have been
 observed in HI and in H$\alpha$ with pointed observations (listed in GOLDmine).
 Black circles mark the position of the seven groups found in region 2. 
 Bottom panel: recessional velocity versus R.A. distribution of 2416 galaxies in the
 region 2 covered by ALFALFA (black).
 683 galaxies detected by ALFALFA are represented in blue. The 383 galaxies observed in H$\alpha$
 as part of H$\alpha3$ are plotted in red. Note the filamentary structure departing from the Coma cluster
 (identified by the "finger of God" near R.A.=195), with increasing velocity toward the Hercules supercluster 
 (R.A.= 240).}
 \label{campione} 
\end{figure*}

\section{The Sample}
\label{sample}

The present analysis is focused on galaxies members of the Coma supercluster,
which belongs to three subsamples: sample 1 is composed of 5026 optically selected 
galaxies brighter than $r\leq$17.77 (and is complete to this limit), sample 2 consists of 683
HI-selected galaxies (from ALFALFA), sample 3 (a subset of sample 2)
consists of 383 galaxies that were followed-up with H$\alpha$
observations as part of H$\alpha3$.

The sky distribution of the Great Wall
is displayed in the top panel of Figure~\ref{campione}, while a wedge diagram
of the HI-selected sample is shown in the bottom panel.  The optically
selected sample was built from the union of two regions: region 1 with $\rm
172^o < R.A. < 202^o\, ; \,18^o< Dec. <32^o$ ($\rm 11.5^h <R.A. <13.5^h$, gray
shaded in Figure \ref{campione}), which traditionally describes the Coma
supercluster, and region 2 with $\rm 150^o < R.A. < 240^o\, ; \,24^o< Dec. <28^o$
($\rm 10^h <R.A. <16^h$, blue-shaded in Figure \ref{campione}) covered by
ALFALFA.  In both regions we limit our analysis to the redshift interval
$3900<cz<9500~{\rm km~s^{-1}}$, chosen to comprise the ``finger of God'' of the
Coma cluster.

\subsection{Optically selected sample}

Galaxies belonging to the Coma supercluster, which lies at a mean distance of 100 Mpc, have
apparent radii small enough that their SDSS images are only little affected by the ``shredding" of large galaxies
into multiple pieces, which leads to wrong magnitude determinations\footnote{Shredding of large galaxies 
is caused by the combined effect of cutting the 
images of individual galaxies into multiple pieces, due to the observing strategy of SDSS by discrete ``tiles'', and to 
the inefficiency of Sextractor (Bertin \& Arnouts 1996)
in reconstructing the total flux of very extended objects.} (Blanton et al. 2005a,b,c). 
Hence 
a reliable optically selected sample can be extracted from the SDSS spectroscopic database 
following the criteria described in Gavazzi et al. (2010, 2011) 
which we only briefly summarize here.
We searched the SDSS DR7 spectroscopic database (Abazajian et al. 2009) in regions 1 and 2 (see Figure~\ref{campione})
 for all galaxies 
with $r\leq$17.77 mag, which matches the selection criterion of the SDSS spectral catalog (Strauss et al. 2002), in the redshift interval $3900<cz<9500~{\rm km~s^{-1}}$. 
We obtained 4790 targets. For each, we extracted the coordinates, the
$u,g,r,i,z$ Petrosian magnitudes (AB system, not corrected for internal extinction), and the spectroscopic information, 
including the principal (nuclear) line intensities and the redshift. 
The morphological classification of all galaxies was performed by individual visual inspection of
SDSS color images.

To fill in the incompleteness of SDSS for luminous galaxies due to shredding and fiber conflict, 
we added 133 CGCG (Zwicky et al. 1961-1968)
galaxies with known redshifts from NED that were not included in the SDSS spectral database.  
For these objects, we also took the $u,g,r,i,z$ Petrosian magnitudes using the SDSS DR7 navigation tool,
which provides accurate magnitudes. 
Additional galaxies that could not be found in the SDSS spectroscopic catalog were searched for using NED.
For these targets, we again evaluated 
the $u,g,r,i,z$ magnitudes using the SDSS navigator, and of these we selected 76 objects that meet 
the condition $r\leq17.77~\rm mag$.
We repeated a similar search in the ALFALFA database (Haynes et al. 2011) in region 2,
where we found 28 additional HI-selected systems with $r\leq17.77~\rm mag$, that were not included in the SDSS spectral database. 

In total, our optical sample consists of 5026 galaxies: 4790 from SDSS and 236 from other sources.
Of the 5026, 2146 galaxies lie in region 2. 

The stellar mass was derived from the $i$ magnitudes and $g-i$ color using the transformation 
\begin{equation}
\log \Bigl(\frac{M_{*}}{M_{\odot}} \Bigr) = -1.94 + 0.59 \cdot (g-i) + 1.15 \cdot \log \Bigl(\frac{L_i}{\rm L_{\odot}} \Bigr), 
\label{eq:our_mass}
\end{equation} 
where $\log L_{i}$ is the $i$ band luminosity in solar units
($\log L_{i}=(I-4.56)/-2.5$). This is a modification of the Bell et al. (2003)
formula that we computed to be consistent with the mass determination of Brinchmann MPA-JHU\footnote{www.mpa-garching.mpg.de/SDSS/DR7/, see Salim et al. (2007)}.

Following a  procedure identical to the one used in Gavazzi et al. (2010, 2011), the local number density 
$\rho$ around each galaxy was computed within a cylinder 
of 1 $h^{-1}~\rm Mpc$ radius and 1000 $\rm km~s^{-1}$  half-length.
Around each galaxy we computed the 3-D density contrast as 
$$\delta_{1,1000} = \frac{\rho-<\rho>}{<\rho>}$$,
where $\rho$ is the local number density and $<\rho>$ = 0.05 gal $(h^{-1}~$Mpc$)^{-3}$ 
represents the mean number density measured in the whole region. 
We divided the sample into four overdensity bins, chosen to highlight physically different
environments of increasing level of aggregation:
the ultra-low density bin (UL: log(1+$\delta_{1,1000})\leq 0$) that describes the underdense cosmic structures (sparse filaments and voids);
the low-density bin (L: $0 <\rm log(1+\delta_{1,1000}) \leq 0.7$) that comprises the filaments in the Great Wall and the
loose groups; the high-density bin (H: $0.7 <\rm log(1+\delta_{1,1000}) \leq 1.3$) that includes
the cluster outskirts and the significant groups; and the ultra-high density bin (UH: log$(1+\delta_{1,1000}) > 1.3$) 
that corresponds to the cores of rich clusters. 

\subsection{HI-selected sample}
\label{radioselection}

The HI-selected sample analyzed in this work is drawn from the 360 square degree region 2.
This region has been fully mapped by ALFALFA, providing us with a complete sample 
of HI-selected galaxies (Haynes et al. 2011).  
At present, half of the Coma cluster lies within the footprint of ALFALFA.

The completeness and sensitivity of ALFALFA are clearly described and discussed
  in detail in  Saintonge (2007), Martin et al. (2010), and Haynes et al. (2011).
An estimate of the limiting sensitivity of ALFALFA at the distance of Coma can be obtained
using equation 1 in Giovanelli et al. (2007): 
\begin{equation}
S/N=\left(\frac{1000 \cdot S_{21}}{W_{50}}\right) \cdot  \frac{w_{\rm smo}^{1/2}}{\sigma_{\rm rms}},
\end{equation}
where $S_{21}$  is the integrated flux under the HI 
profile in units of Jy km s$^{-1}$, $w_{\rm smo}$  is a smoothing width expressed as the number of spectral resolution bins of 10 $\rm km~s^{-1}$  
that bridges half of the signal width. 
$W_{50}$ is the profile width at 50\% of the height and 
$\sigma_{\rm rms}$ is the typical noise in the baseline in bins of 10 $\rm km~s^{-1}$.
For $W_{50}< 400~\rm km~s^{-1}$, $w_{\rm smo}=W_{50}/(2 \cdot 10)$. 
The HI mass is
$M_{\rm HI}= 2.36 \cdot 10^5 \cdot  S_{21} \cdot D^2$,  
where $D$ is the distance to the source in Mpc.\\ 
The limiting HI mass of ALFALFA can be computed as
\begin{equation}
M_{\rm HI,lim}= 2.36 \cdot 10^5 \cdot D^2 \cdot (W_{50} \cdot 20)^{1/2} \cdot \sigma_{\rm rms} \cdot S/N_{\rm lim},
\end{equation}
where
$W_{50} = W_{50,0} \cdot \sin(incl)$ 
($W_{50,0}$ is the face-on line width)
and $incl$ is the galaxy inclination in the plane of sky determined from equation (A1). 

Adopting $\sigma_{\rm rms}=2.1$ mJy, typical of ALFALFA spectra, the  mean distance to Coma $D=100~\rm Mpc$, 
$S/N_{\rm lim}=6.5$ and $W_{50,0}=100~\rm km~s^{-1}$, we obtain $\log(M_{\rm HI,lim}/{\rm M_\odot})=8.78, 9.08, 9.15$ 
for $incl$=10, 45, 70 deg\footnote{At the distance of Virgo ($D$=17 Mpc) these limits are 35 times lower, becoming 
7.24, 7.54, 7.60 $\rm M_\odot$. }.
In conclusion, ALFALFA is sensitive to galaxies that contain approximately $\log
(M_{\rm HI}/{\rm M_{\odot}}) \sim 9$ at the distance of Coma.

\begin{figure} 
 \centering
\includegraphics[width=\columnwidth, trim = 0cm 0cm 0.8cm 0cm]{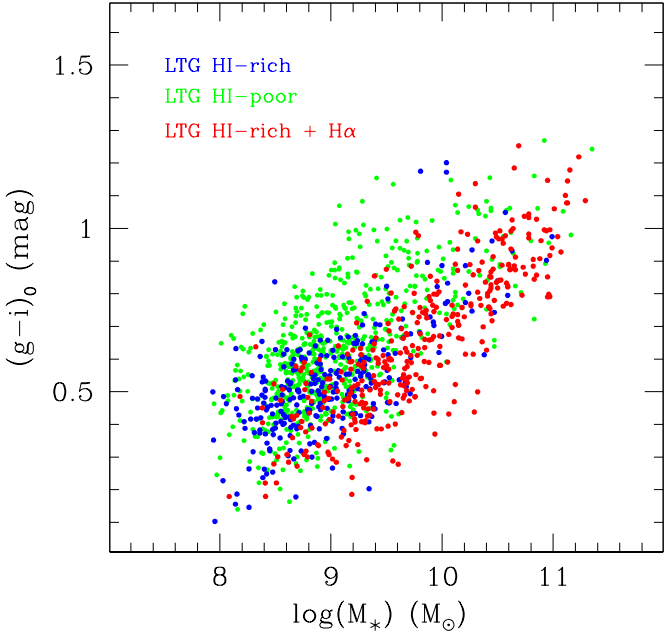}
\caption{Corrected color ($g-i)_o$ vs. stellar mass diagram of optically selected galaxies limited to LTGs (subsample 2). 
The colors are corrected for extinction in the Milky Way and for internal extinction according to the prescriptions
of Appendix A. LTGs  undetected by ALFALFA are represented by green symbols.  LTGs detected by ALFALFA are given in blue. 
 Red symbols denote the subsample of the ALFALFA-selected galaxies that were followed-up in H$\alpha3$.
 }
\label{colmagHa} 
\end{figure}

\subsection{H$\alpha$ sample}
\label{Haselection}

H$\alpha3$ consists of follow-up H$\alpha$ observations of 383 galaxies among
the HI-selected (detected) ones.  These data, resulting from the joined
effort by the H$\alpha3$ collaboration in 2010, 2011 and 2012 using the 1.5m
and 2.1m telescopes at SPM, will be discussed in full detail (and publicly
released) in Paper V of this series.  Starting from 683 ALFALFA targets with
high signal-to-noise (typically S/N$~>6.5$) and a good match between two independent
polarizations (code = 1 sources; Giovanelli et al. 2005, Haynes et al. 2011),
in the first two years (2010, 2011) our observational effort had been focused
on radio targets brighter than $1 ~\rm Jy~km~s^{-1}$ and with $3900<cz<9000~{\rm
  km~s^{-1}}$, while fainter targets were observed in 2012.  However,
approximately 95 additional bright CGCG galaxies (Zwicky et al. 1961-1968) in
the intersection between region 1 and 2 ($\rm 11.5^h <R.A. <13.5^h\, ; \,24^o<
Dec. <28^o$) and  207 additional CGCG galaxies outside region 2 were already
observed in H$\alpha$ by Gavazzi et al. (1998, 2002a, 2002b, 2006), Boselli \&
Gavazzi (2002), and  Iglesias Paramo et al. (2002).  For these targets images and
fluxes are publicly available via the GOLDmine web server (Gavazzi et
al. 2003).  In total, the number of galaxies in Sample 3 (H$\alpha3$) is 383,
but the total number of galaxies observed in H$\alpha$ is 95+383+207=685.
For these targets the SFR was computed from the luminosity of the H$\alpha$
line after correcting for Galactic extinction and after the  H$\alpha$ line was 
deblended from the contribution from [NII] emission using the criteria adopted in Paper I. 
Based on the mass-metallicity relation (Tremonti et al. 2004), we calibrated 
a relation between the ratio N[II]/H$\alpha$ and the absolute $i$ band magnitude, excluding 
active galactic nuclei (AGNs).  
Details
on these quantities will be given in Paper V.
No correction for internal dust extinction has been applied because very few  
drift-scan-spectra are available from Gavazzi et al. (2004) and Boselli et al. (2013), 
which are necessary to estimate the Balmer decrement and in turn
the extinction coefficient A(H$\alpha$) on the full disk scale. 
The SDSS spectra could not provide for this correction because, at the distance of Coma, 
they cover just the nuclear galaxy region
and hence are not representative of the whole galaxy.

Summarizing, ALFALFA  
provides us with the HI mass of LTGs complete to approximately
$10^9~\rm M_\odot$, whose optical colors (corrected for internal extinction
according to the prescriptions of Appendix A) and stellar masses are
distributed as shown in Figure \ref{colmagHa}.  
The figure highlights that  only the part of the ``blue cloud''
composed of HI-rich LTGs is sampled by ALFALFA (blue + red symbols), whereas
LTGs that have less than $10^9~\rm M_\odot$ of HI, hereafter considered HI-poor
LTGs, are not detected by ALFALFA (green symbols). 
in this Figure we highlight with red symbols We highlight the 
subsample of ALFALFA targets that were followed-up in H$\alpha3$ with red symbols.
         \begin{figure}
	 \centering
	 \includegraphics[width=\columnwidth, trim = 1cm 1cm 0cm 0.5cm]{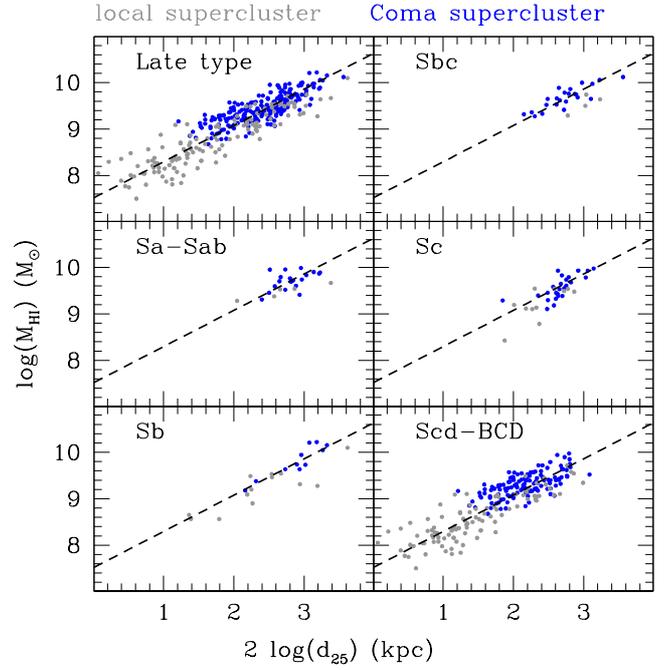}
	 \caption{$\log M_{HI}(T,D_{25})$ relation for isolated galaxies with $\delta_{1,1000}<0$ in the Coma supercluster
	 (blue) and in the Local Supercluster (Paper II; gray). 
	 The fit obtained using all late-type (Sa-BCD) objects is given as a dashed line
	 in every panel.}
	 \label{MHIdiam} 
	 \end{figure}

 	\begin{figure}
	\centering
	 \includegraphics[width=\columnwidth, trim = 1cm 1cm 1cm 0cm]{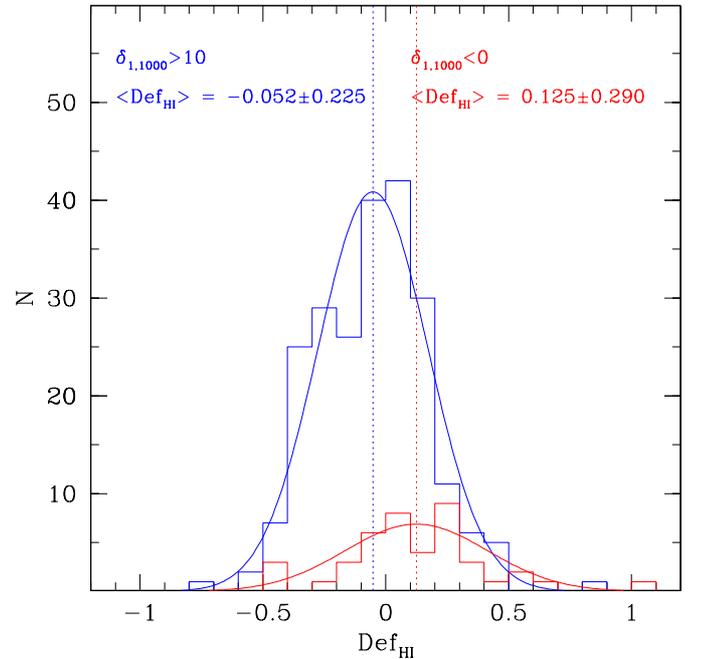}
       \caption{Histogram of the $Def_{HI}$ parameter derived from ALFALFA observations of the Coma supercluster, separately 
	 for isolated  objects with $\delta_{1,1000}<0$ (blue) and for galaxies in the densest regimes 
	($\delta_{1,1000}>10$  red).
	 }
	\label{defhist} 
	\end{figure}
	\begin{figure}
	\centering
	\includegraphics[width=\columnwidth, trim = 0.5cm 1cm 0.8cm 0cm]{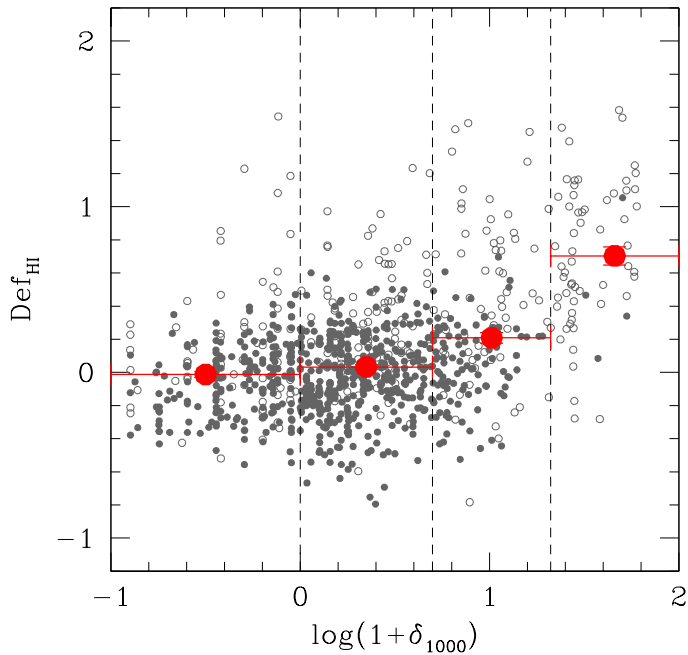}
       \caption{Distribution of the $Def_{HI}$ parameter of galaxies in the Coma supercluster as a function
       of the local galaxy density $\delta_{1,1000}$.  Filled dots represent HI data from ALFALFA, empty symbols derive from
       deeper, pointed HI measurements of optically selected galaxies. Clearly ALFALFA is not sufficiently
       deep to detect deficient objects, except for few massive, face-on galaxies (the only point above $Def_{HI}$=1
       is NGC 4921, a face-on giant Sb in the Coma cluster). Deeper, pointed HI observations reveal
       the expected increase of the mean HI deficiency with increasing local galaxy density (red symbols). 
	 }
	\label{sboro} 
	\end{figure}
\section{Environmental Effects}
\label{env}
\subsection{Pattern of HI-deficiency}
\label{hidef}

It has long been known that the HI-deficiency parameter  ($Def_{\rm HI}$) is perhaps the most reliable indicator
of whether or not an LTG belongs to the harsh environment of a rich cluster (Giovanelli \& Haynes, 1985).
$Def_{\rm HI}$ has been defined by Haynes \& Giovanelli (1984)
as the logarithmic difference between the HI mass observed in an object 
and the expected value in isolated and unperturbed  objects of 
similar morphological type $T$ and linear diameter $D_{25}$:
$Def_{\rm HI}=\langle \log M_{\rm HI}(T,D_{25})\rangle- \log M_{\rm HI}^{\rm obs}$. Here, 
$\langle \log M_{\rm HI}(T,D_{25})\rangle= C_1+C_2\times 2 \log D_{25}$,
where $D_{25}$ (in kpc) is determined in the $g$ band at the $25^{th}$ $\rm mag ~arcsec^{-2}$ isophote.
The coefficients $C_1$ and $C_2$ were determined by 
Haynes \& Giovanelli (1984) by studying a control sample of isolated objects, 
and later, on a larger sample by  Solanes et al. (1996).   
Both samples are composed almost exclusively of giant spirals however.
The HI deficiency parameter was therefore poorly calibrated for dwarf objects.
This problem was addressed in Paper II, where the relatively isolated galaxies in the Local Supercluster 
were used to determine that $C_1$=7.51 and $C_2$=0.68 hold for all LTGs.
Here, combining isolated LTGs from the Local Supercluster with the lowest density objects in the Coma supercluster we obtain
$C_1$=7.50 and $C_2$=0.79, consistently with the Local Supercluster alone (and with the Coma1 cloud, 
see Boselli \& Gavazzi 2009).
These coefficients are consistent with those adopted by Solanes et al (1996) up to Sc galaxies, and  
by Toribio et al. (2011) up to Scd-Sd galaxies.

Figure \ref{MHIdiam} shows the relations among $\rm M_{HI}$ and the diameters found adopting these coefficients. 
The fit\footnote{Linear regressions are obtained in this work using the 
bisector method (mean coefficients of the direct and the inverse relation, Isobe et al. 1990).} obtained for all LTGs 
is depicted with the dashed line.
The distribution of the $Def_{\rm HI}$ parameter of galaxies in the Coma supercluster 
is plotted in Figure \ref{defhist} separately for 
$\delta_{1,1000}<0$ (isolated objects, blue) and $\delta_{1,1000}>10$ (rich clusters, red), with the mean
$Def_{\rm HI}$ obtained in the two bins 
(intermediate-density objects are not plotted, but are considered in the following analysis). 
Apparently, isolated objects have a mean deficiency consistent with $0 \pm 0.2$. 
Throughout  this paper we consider as ``normal'' or ``unperturbed'' by the cluster environment, 
as far as their HI content is concerned, 270 galaxies 
with $Def_{\rm HI} \leq 0.2$, i.e., within 1 $\sigma$ of the mean deficiency of the isolated sample. 
At the distance of Coma, galaxies with $Def_{\rm HI}> 0.2$ are undersampled because they lie below the 
sensitivity of ALFALFA.

Figure \ref{sboro} shows the deficiency parameter plotted as a function of the local density.
Filled symbols refer to galaxies in the HI-selected sample, i.e., detected by ALFALFA. Evidently,
as the galaxy density $\log(1+\delta_{1,1000})$ approaches 1 (the cluster outskirts) ALFALFA detects almost no galaxies,  
because they are too shallow to reveal the Giovanelli \& Haynes (1985) pattern of HI deficiency near the Coma cluster.   
However, when pointed long-integration HI observations deeper than ALFALFA are considered, such as those available for region 1 from GOLDmine
(open gray symbols with red binned averages), the expected increase of $Def_{\rm HI}$ with local density clearly appears.

\subsection{What do the LTGs undetected by ALFALFA represent?}
\label{undet}
 
 In contrast to the Local Supercluster (Paper II of this series),
 the sensitivity of ALFALFA at Coma is insufficient for detecting galaxies with significant HI deficiency 
 (see previous section)\footnote{We recall that ALFALFA is a two-passes survey of approximately 7000 sq degrees visible from Arecibo.
To make it sensitive to $10^8 ~ \rm M_\odot$ of HI at the distance of Coma, it would require some 500,000
hours of telescope time!}.
 The known trend of increasing HI deficiency with increasing galaxy density
 (see Figure \ref{defhist}) could  be revealed only
 using the long-integration pointed radio detections available in region 1.
 Therefore a discussion similar to Paper II, where some galaxy properties were 
 analyzed in bins of increasing $Def_{\rm HI}$, cannot be undertaken. 
 Moreover, because H$\alpha3$ suffers from the same selection bias as ALFALFA, 
 no trends between the SFR and the HI deficiency  can be directly investigated.\\  
 This problem can be addressed, however,
 by separating LTGs that were detected/undetected by ALFALFA.
\begin{figure}
 \centering
  \includegraphics[width=\columnwidth, trim = 0.5cm 1cm 0.8cm 0cm]{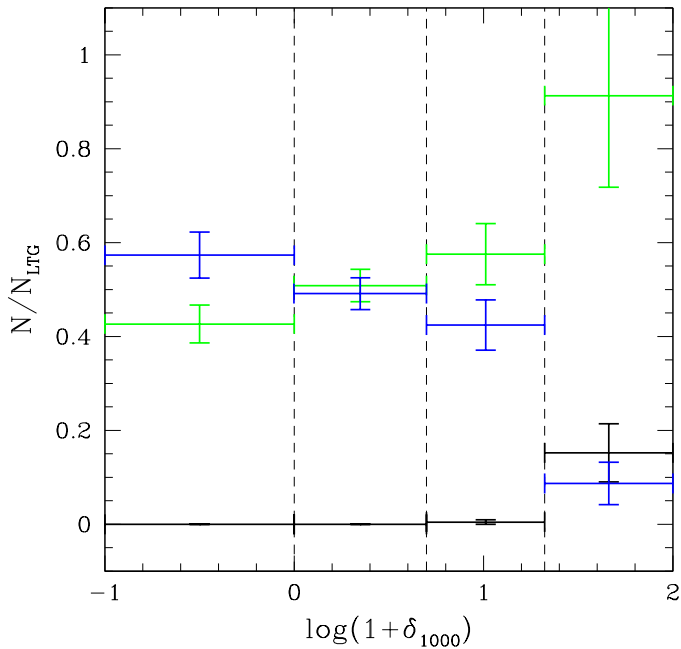}
  \caption{The fraction of LTGs (in region 2) 
detected by ALFALFA (blue histogram); undetected by ALFALFA (green histogram);  
PSB (black histogram),  in four bins of increasing local density 
(representing the sparse cosmic web, the filaments, the groups and the cluster's cores). }
 \label{density} 
 \end{figure}
 As already stressed in Paper I, the radio selection biases H$\alpha3$  toward HI-rich 
LTGs (see also Gavazzi et al. 2008).
ALFALFA detects almost exclusively LTGs in the blue sequence, not all LTGs are detected 
(see green symbols in Figure \ref{colmagHa}). 
As we showed in Paper II and in Gavazzi et al. (2008), where we focused on galaxies in the Local Supercluster,
the very existence of a limiting HI mass combined with the 
relation $\log M_{\rm HI}=0.52 \times \log M_*+4.54$ (see Appendix B)
that holds for normal galaxies 
induces an optical limit for the detection of HI-rich systems. At the distance of Coma, 
galaxies fainter than $\log(M_*/$M$_{\odot})\sim 8.5$  or $M_i \sim -17$ mag are not detected by ALFALFA. 
Furthermore, even some late-type galaxies brighter than this limit are not detected in ALFALFA 
or in H$\alpha3$. 
LTGs that are not detected by ALFALFA are truly HI-poor systems\footnote{Before assuming that all galaxies undetected by ALFALFA are
  intrinsically HI-poor, i.e.,  have $\log (M_{\rm HI}/$M$_{\odot})<9$, one must
  exclude the optical targets undetected by ALFALFA due to confusion in the
  Arecibo beam by other detected galaxies at similar velocity.  Using a 3.5
  arcmin FWHM beam and $\Delta V=200~\rm km~s^{-1}$ at the position of each
  detected target, we found only 28 optically selected candidates that were
  confused (roughly 1\% of all galaxies in region 2 and 4\% of all
  undetected LTGs). We thus conclude that most undetected LTGs with  $\log(M_*/$M$_{\odot})> 9$ are truly
  HI-poor systems. Indeed, using pointed observations available from Arecibo
  for LTGs in the intersection between region 1 and region 2, we determined that
  $<Def_{\rm HI}>=0.13$ for LTGs detected by ALFALFA, while $<Def_{\rm HI}>=0.73$ for
  LTGs undetected by ALFALFA.}.

Figure \ref{density} sheds more light on the nature of HI-poor LTGs by plotting their 
frequency distribution as a function of the local galaxy density:
the fraction of HI-rich LTGs detected by ALFALFA (blue histogram), representing about 50\% of all LTGs in the field, 
decreases gradually with increasing density and  drops to zero in the core of Coma; 
the fraction of HI-poor LTGs undetected by ALFALFA 
(green histogram), representing the remaining 50\% of all LTGs in the field, 
increases with increasing density to nearly 100\%. Incidentally, 
the fraction of post-star burst systems is null in the field, in filaments, and in groups but it becomes significant in the core of Coma
(see also Gavazzi et al. 2010).
Summarizing, when galaxies approach denser environments, 
their gas content is progressively reduced, which mimics the pattern of morphology segregation.

\subsection{Color, HI content, and local density}
\label {HIcontent}

 \begin{figure*}
 \centering
\includegraphics[width=7.5cm,height=7.5cm]{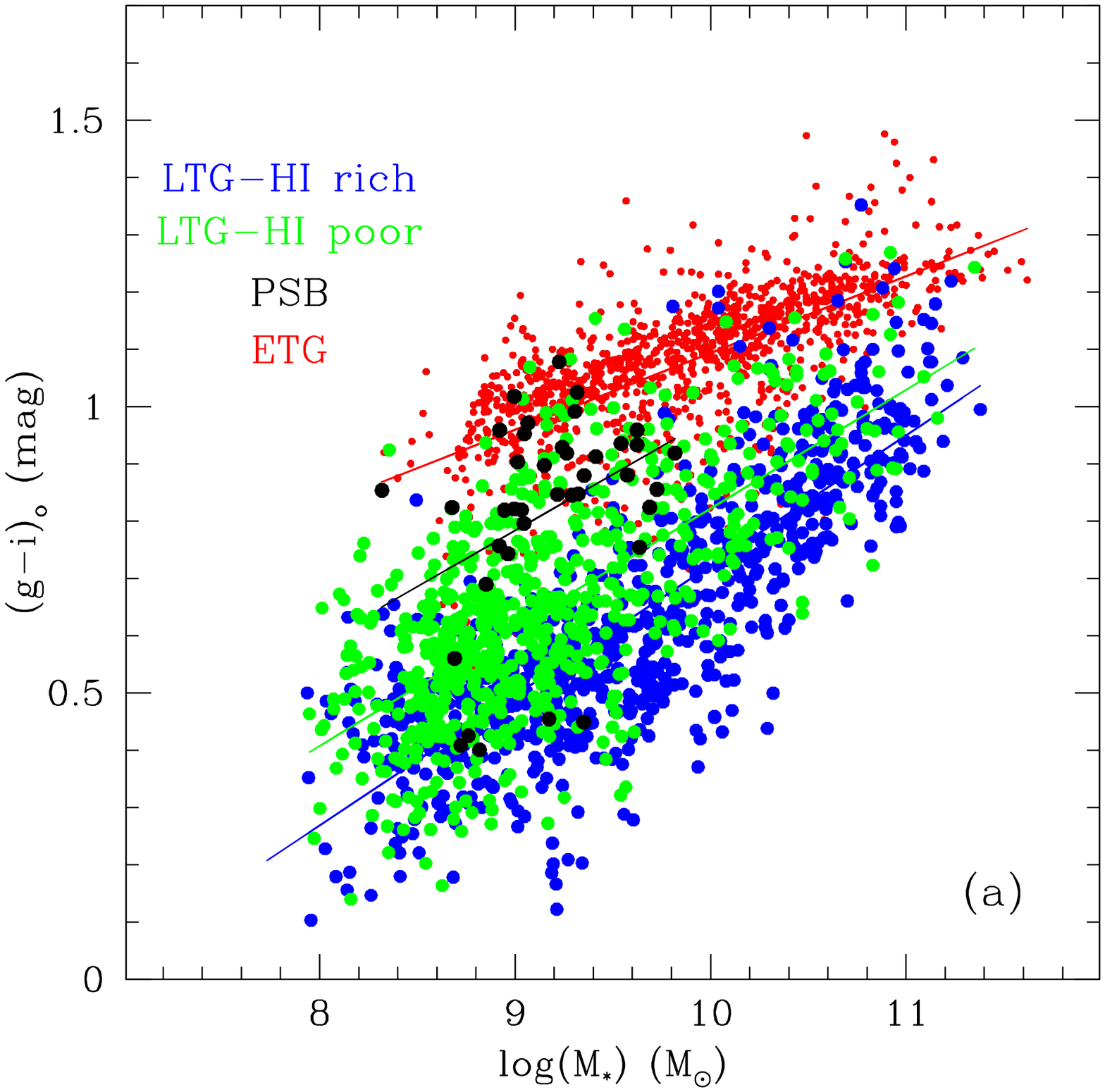}
\includegraphics[width=7.5cm]{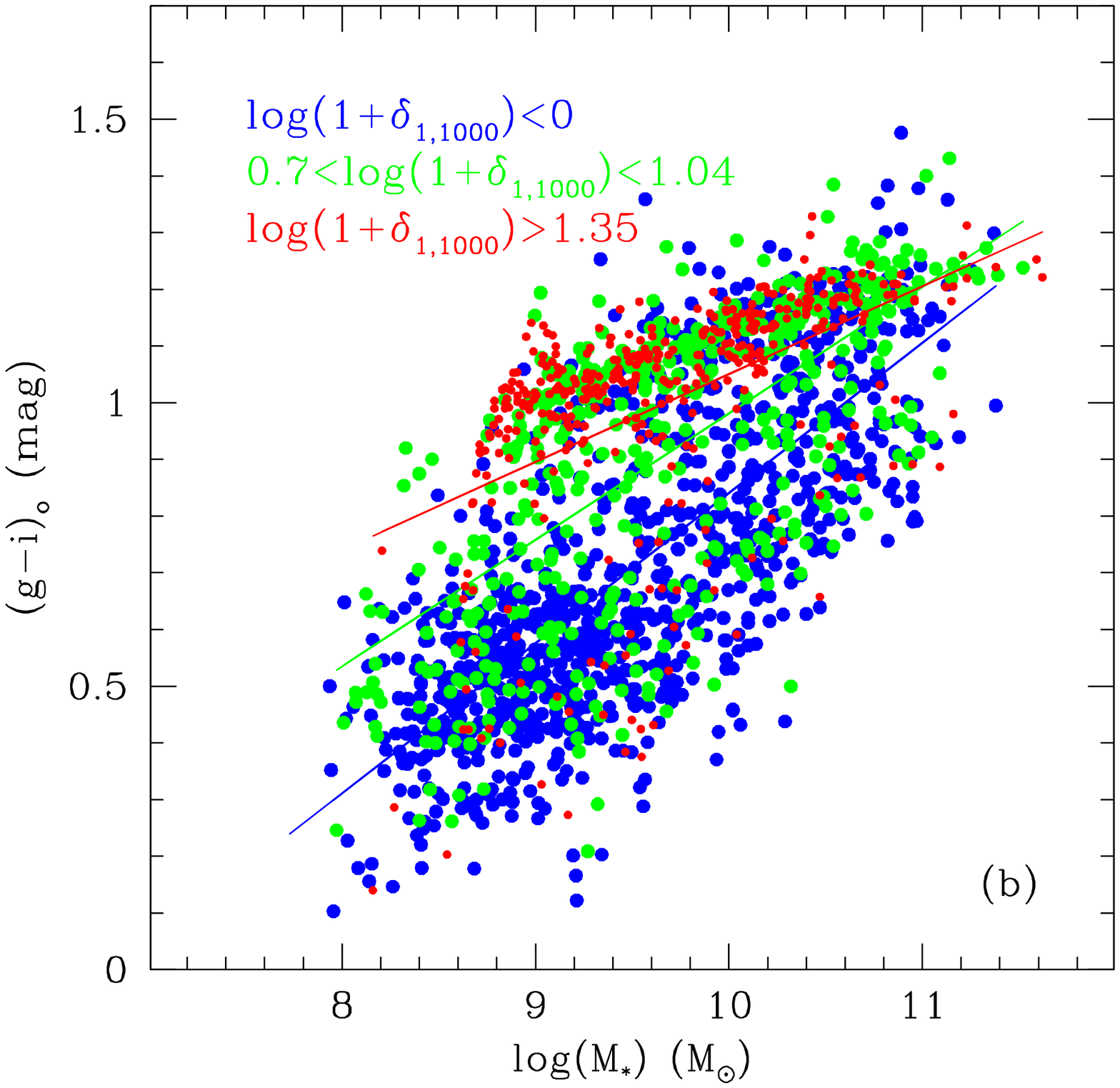}
\includegraphics[width=7.5cm]{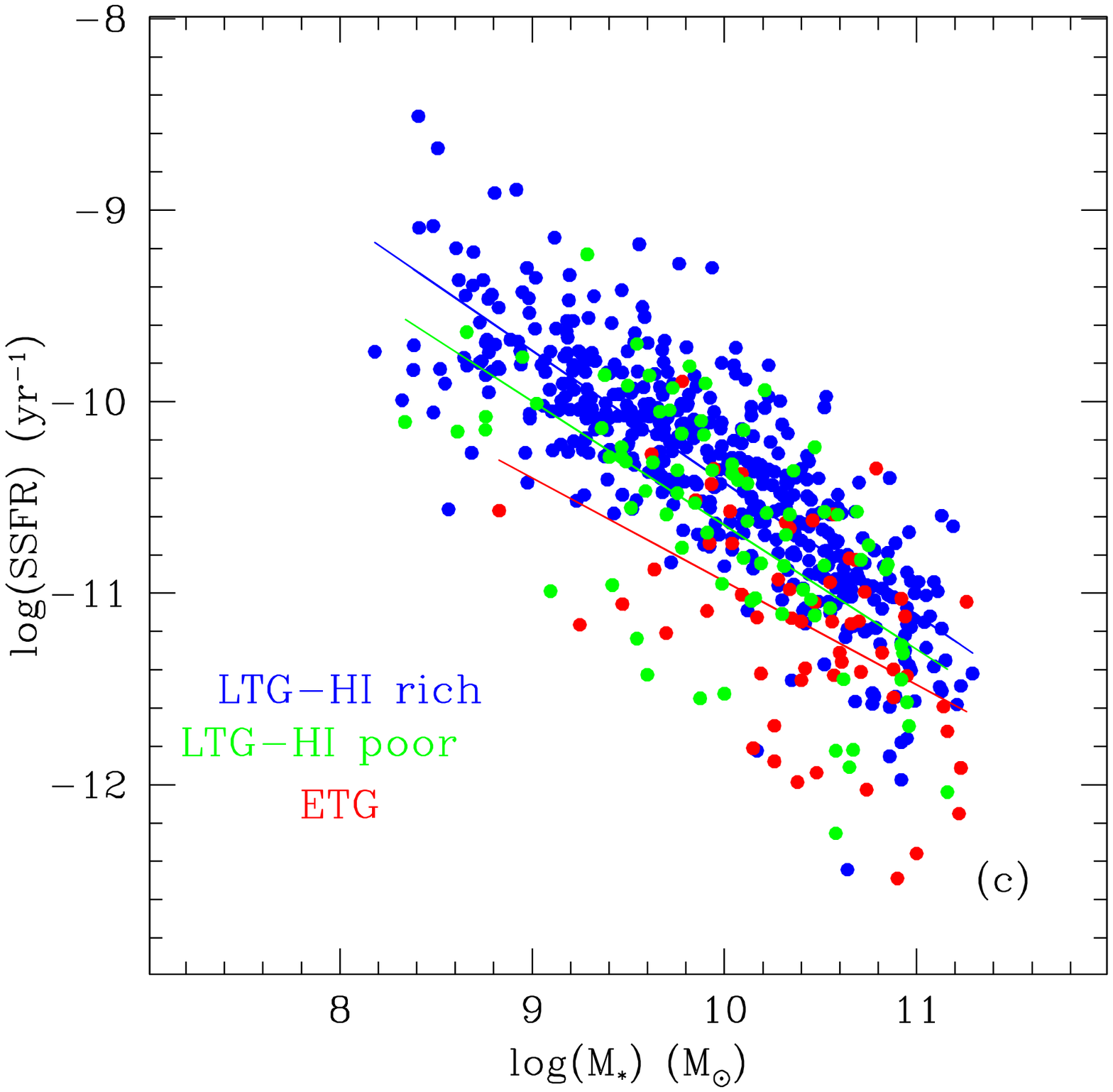}
\caption{
  Four-step sequence of galaxies of increasing $(g-i)_o$ color is
  identified in the color-mass diagram of panel (a) when four groups of
  different morphological types/gas-content/nuclear properties are considered.
  Panel (a): corrected color ($g-i)_o$ vs. stellar mass diagram of 
  optically selected galaxies  of all morphological types. 
  Red symbols represent ETGs.  Green symbols represent HI-poor
  LTGs not detected by ALFALFA, while HI-rich LTGs detected by ALFALFA are
  given in blue. Large black symbols represent PSB (also named k+a, i.e., with Balmer lines in absorption) galaxies identified by
  Gavazzi et al. (2010).  They all lie at $\log (M_*/{M_{\odot}})< 10$. 
  Panel (b): a similar gradual trend in the
  color-mass diagram is found when galaxies are grouped in bins of increasing local
  density. This panel shows the corrected color ($g-i)_o$  vs. stellar mass diagram
  of optically selected galaxies  of all morphological  types,
  in three bins of decreasing local density: red=   $\log(1+\delta_{1,1000})>1.35$, 
  green=  $0.7<\log(1+\delta_{1,1000})<1.04$; blue=   $\log(1+\delta_{1,1000})<0$. 
  Panel (c):  the specific star formation rate (SSFR) vs. stellar mass diagram of
  galaxies covered by ALFALFA and followed-up by
  H$\alpha3$. Galaxies are color-coded according to the following
  criterion: red = ETGs; green = LTGs undetected by ALFALFA (HI-poor);
  blue = LTGs detected by ALFALFA (HI-rich).  Unsurprisingly, the statistics are very poor
  among ETGs and for HI-poor LTGs.
  }
\label{colmagaa} 
\end{figure*}

We now show that HI-poor LTGs are statistically less star-forming than their HI-rich
counterparts of similar stellar mass. 
Figure \ref{colmagaa}a shows the color-mass relation of galaxies, divided into
 four groups of different morphological type/gas-content: HI-rich (detected by ALFALFA) LTGs are plotted with blue symbols; 
HI-poor (undetected by ALFALFA) LTGs are marked with green symbols,
PSB galaxies identified by Gavazzi et al. (2010)  are shown with large black symbols\footnote{A confirmation 
that PSB galaxies are low-mass objects that fill the
``green valley''  between star-forming   and passive galaxies was derived for the galaxy Zoo by Wong et al. (2012).}, and
red symbols represent early-type galaxies that belong to the red sequence.
  
Late-type galaxies that are not detected by ALFALFA (green symbols), dubbed HI-poor LTGs, 
 lie nearer to the ``green valley'' than ALFALFA detected galaxies 
(blue symbols) even though they belong to the sequence of LTGs. 
For all types of galaxies, Figure \ref{colmagaa}a reports the
linear fit of the color-stellar mass relation. 
The four galaxy groups, selected by their
optical morphology/gas content and nuclear spectral properties, have
increasingly redder colors, i.e., decreasing SSFR.  
Despite the paucity of H$\alpha$ measurements (particularly of HI-poor and ETGs),
Figure \ref{colmagaa}c
supports the idea that the color-mass sequence is indeed a star formation-mass
sequence. Here we plot the specific star formation rate (SSFR), i.e., the star
formation rate per unit stellar mass, as derived from our H$\alpha$ data for
ALFALFA-detected (blue symbols) and for the few HI-poor (green symbols) and
ETGs (red symbols) that are included in H$\alpha3$, with their linear
fits\footnote{By selection we have very few H$\alpha$ images of HI-poor
  galaxies (either ETGs or LTGs) because, as they are not in the ALFALFA sample, they
  are excluded also from H$\alpha3$.  However, a few dozen of them were
  serendipitously observed, some on purpose, during the 2012 run.
  In particular the 80 measurements for ETGs belong for the most part to S0 (19) and S0a (38).
   Their SFR is derived from the intensity of the H$\alpha$ line.}.
\begin{figure*}
\centering
\includegraphics[width=\columnwidth, trim = 0cm 0cm 0.5cm 0cm]{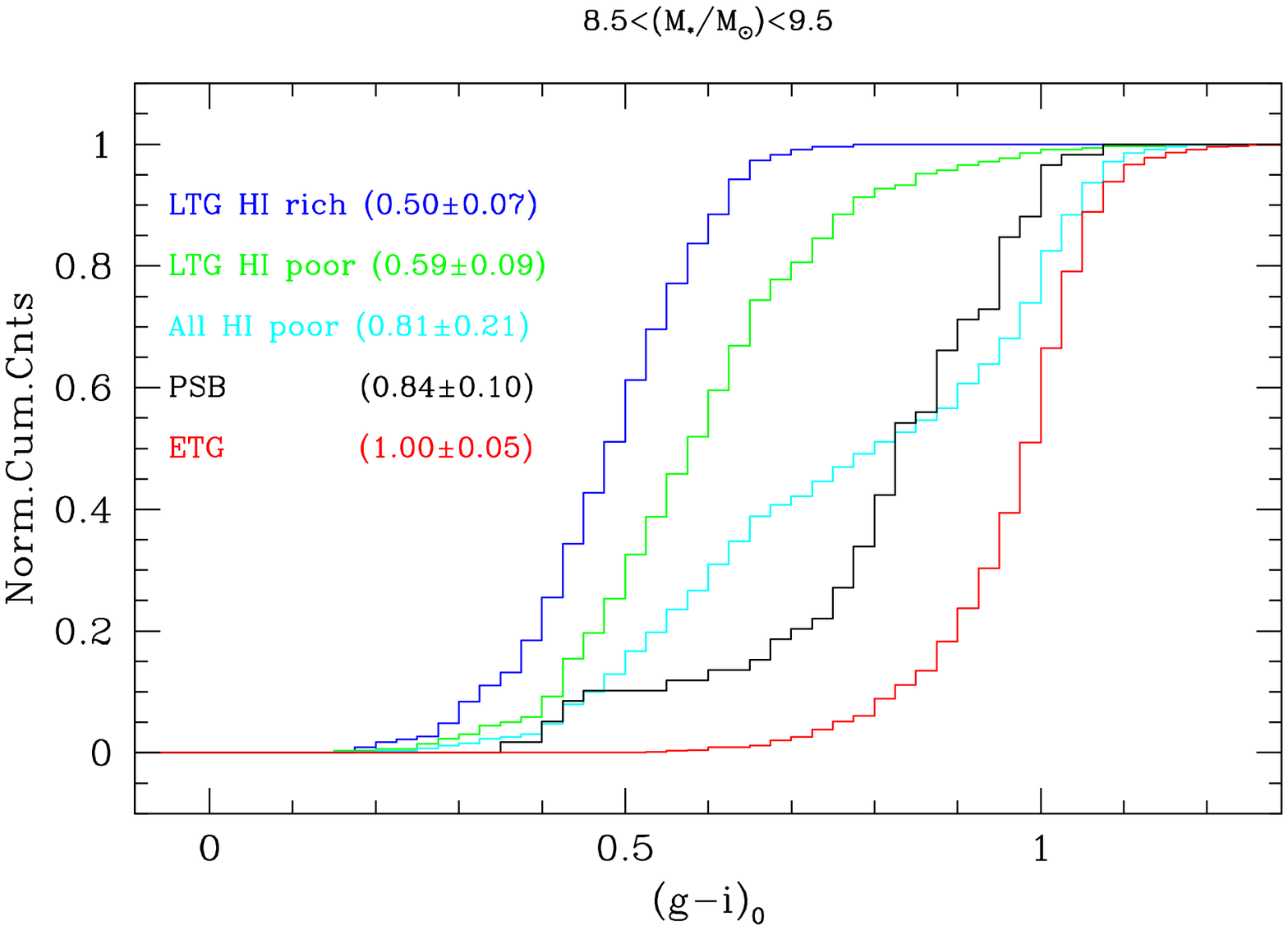}
\includegraphics[width=\columnwidth, trim = 0cm 0cm 0.5cm 0cm]{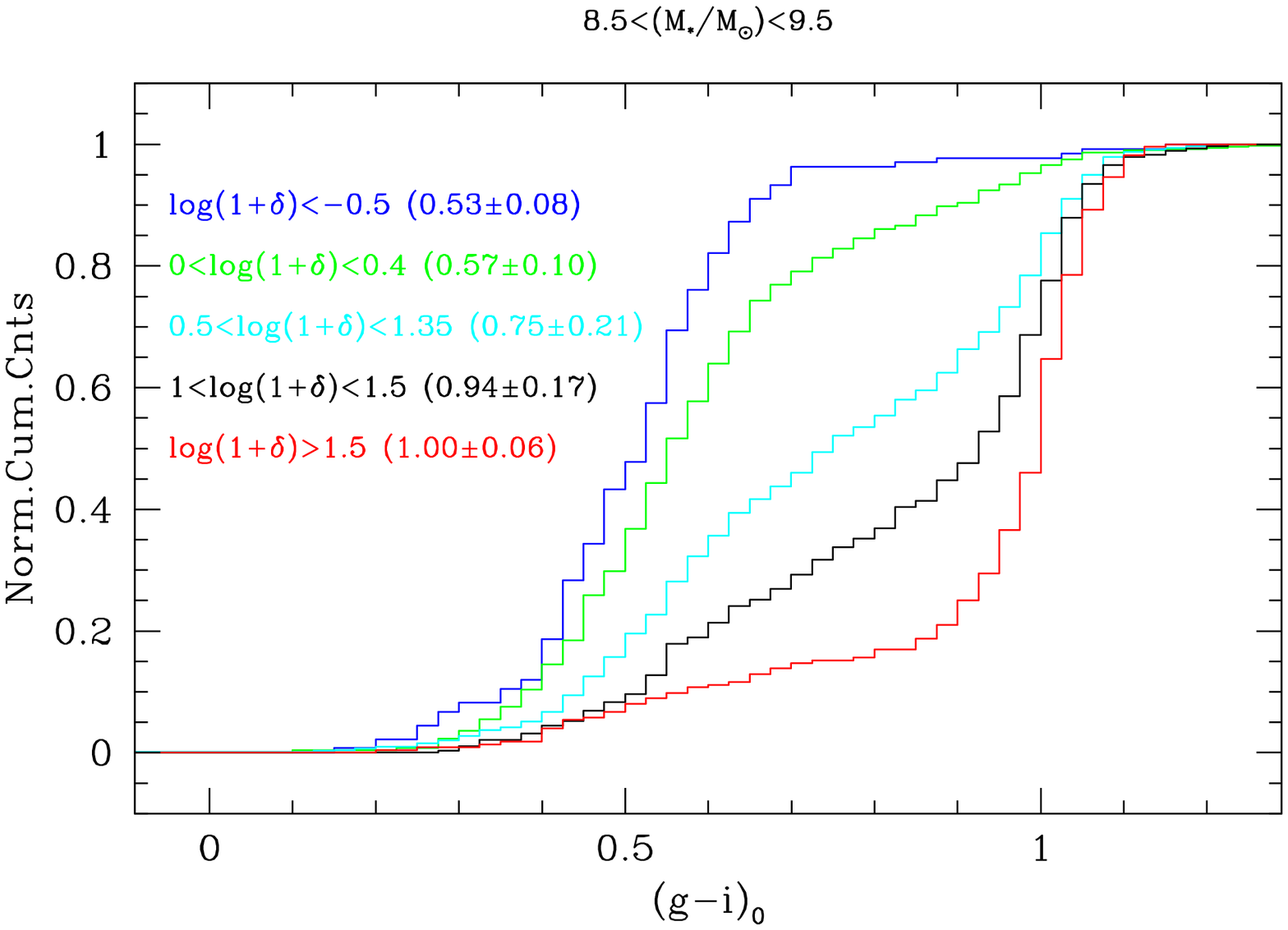}
\caption{Cumulative distributions of the $(g-i)_o$ color in the interval $8.5<\log(M_*/{M_{\odot}})<9.5$, in 
the bins of morphological type/HI content (left panel) defined in Section \ref{HIcontent}
and in bins of local galaxy density (right panel). In parenthesis the median $(g-i)_o$ color with uncertainty is given.
Notice the similarity between the distributions in the two panels. 
}
\label{histo} 
\end{figure*} 

There is a qualitative correspondence (albeit with a large scatter) between the classification given in Fig \ref{colmagaa}a, based on morphology,
and the one given in in Figure \ref{colmagaa}b, based on local galaxy density:
HI-rich LTGs detected by ALFALFA have mean $\delta_{1,1000}=1.7 \pm 4.3$, typical of the sparse galaxy population; 
HI-poor LTGs undetected by ALFALFA have mean $\delta_{1,1000}=4.8 \pm 11$, typical of the outskirts of rich clusters and of the densest groups;
galaxies with nuclear post-star-burst spectra (PSB) and passive ETGs have mean $\delta_{1,1000}=23 \pm 20$,
implying that they are found primarily in the cores of rich clusters.

A quantitative assessment of the median $(g-i)_o$ color in different classes of morphology/HI content and its
correspondence with the median $(g-i)_o$ color  in different classes of local galaxy density is given in Figure \ref{histo},
referring only to the low-mass interval $8.5<\log(M_*/{\rm M_{\odot}})<9.5$.
The left panel shows the cumulative $(g-i)_o$ distribution  of galaxies (in region 2), selected according to a criterion of 
decreasing ``lateness''\footnote{Along with these four categories, Figure \ref{histo} also reports  with cyan symbols the intermediate 
interval that contains HI-poor galaxies of any morphological type for completeness. The median color in this interval is 0.81 mag.}: 
\begin{itemize}
\item{HI-rich LTGs  ($8.5<\log(M_*/{\rm M_{\odot}})<9.5$) (blue)}
\item{HI-poor LTGs  ($8.5<\log(M_*/{\rm M_{\odot}})<9.5$) (green)}
\item{PSB galaxies  ($8.5<\log(M_*/{\rm M_{\odot}})<9.5$) (black)}
\item{ETG galaxies  ($8.5<\log(M_*/{\rm M_{\odot}})<9.5$)  (red),}
\end{itemize}
The median $(g-i)_o$ color increases step by step from 0.50, 0.59, 0.84, to 1.00 mag
along this sequence, which thus represents a sequence of decreasing specific star formation. 
The right panel shows the cumulative $(g-i)_o$ distribution of galaxies of any morphological type,
selected in bins of increasing (not necessarily contiguous) local galaxy density: from $\log(1+\delta_{1,1000})<-0.5$ (blue),
$0<\log(1+\delta_{1,1000})<0.4$ (green), $0.5<\log(1+\delta_{1,1000})<1.35$ (cyan), $1<\log(1+\delta_{1,1000})<1.5$ (black) to
$\log(1+\delta_{1,1000})>1.5$ (red). In this case the median $(g-i)_o$ color increases step by step 
from 0.53, 0.57, 0.75, 0.94, to 1.00 mag,
which agrees well with the previous series, so the shape of the distributions is remarkably similar, 
despite the completely independent criteria in choosing the morphological intervals and the density intervals.
It should be remarked that this result holds for galaxies in the low-mass interval $8.5<\log(M_*/{\rm M_{\odot}})<9.5$ only. 
We have checked that for $\log(M_*/{\rm M_{\odot}})>9.5$ the data do not support a similar pattern with increasing ambient galaxy density.
Consistently with Gavazzi et al. (2010) we conclude that nurture works in the low mass regime, 
while other genetic processes,
namely $downsizing$, have shaped the high-mass part of the red sequence earlier in the cosmic history.
In the high-mass regime Mendel et al. (2012) have suggested that the quenching of the star 
formation, leading to the transition from star-forming to passive galaxies, is driven by the formation 
of stellar bulges.
   
\begin{figure} 
 \centering
\includegraphics[width=\columnwidth, trim = 0cm 0cm 0cm 2.5cm]{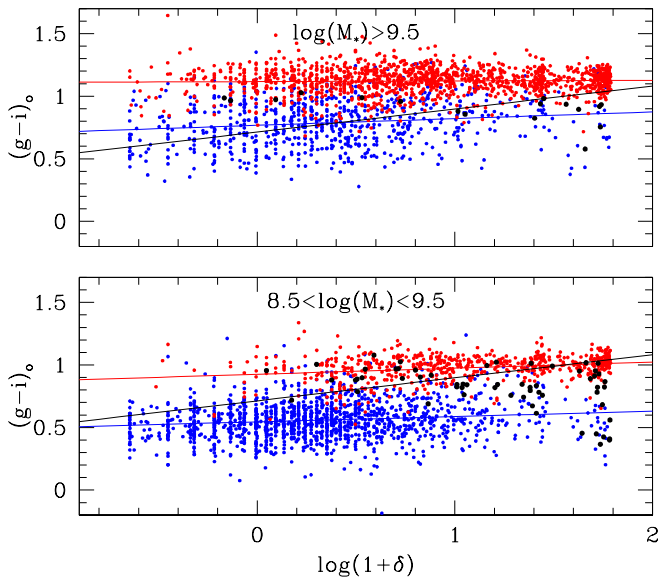}
\caption{Blue and red sequence traced by the $(g-i)_o$ color as a function  of the local galaxy density
in two bins of stellar mass, below (bottom) and above (top) log$(M_*)=9.5$. Blue symbols mark LTGs, while ETGs
are given with red symbols. The two populations are fitted separately (colored lines) and together (black line).
The black points represent PSBs.
 }
\label{coldensity} 
\end{figure}

Figure \ref{coldensity} is useful for discussing the dependence of the $(g-i)_o$ color on the local galaxy density,
separately for galaxies less and more massive than $\log(M_*/{\rm M_{\odot}})=9.5$.
First, the massive galaxies have a well-developed red sequence even at galactic densities as low as $\log(1+\delta_{1,1000})=-0.5$,
while for the less massive objects the red sequence exists only above $\log(1+\delta_{1,1000})>0.5$,
in agreement with Gavazzi et al. (2010).
Second, galaxies in the red sequence dominate in number the distribution of the massive galaxies, while the reverse is true for
low-mass galaxies. 
Third, blue-sequence galaxies are more abundant at low density, where the red sequence is coarser
and the number of PSB galaxies is relevant only at low mass, in agreement with Gavazzi et al. (2010) and
Wong et al. (2012).
Fourth, the color of the ETG sequence is constant over the full density range (see red linear fit), while
the color of the LTG sequence increases only very little by  0.1 mag in the low-mass regime (see blue linear fit).
The apparent slope of the color as a function of density considering all galaxies (see black linear fit) 
is entirely due to a population effect: at high density the ETGs dominate in number. 
It is not clear if the color distribution is truly bimodal or, if we had not subdivided LTGs from ETGs by morphology,
the red and the blue sequence would result in a single sequence that would progressively redden with increasing density.

\subsection{Activity on nuclear scale}

\begin{figure} 
 \centering
\includegraphics[width=\columnwidth, trim = 0cm 0cm 0.5cm 0cm]{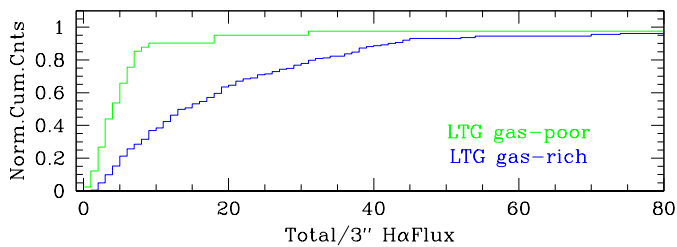}
\caption{
Cumulative distribution of the ratio of total H$\alpha$ flux to H$\alpha$ flux measured in a central 3-arcsec aperture
in the imaging material.
The blue histogram refers to LTGs detected in ALFALFA (HI-rich) and the green one depicts LTGs not detected in ALFALFA (HI-poor).
}
\label{fluxratio} 
\end{figure}

It is interesting to verify if the H$\alpha$ morphology of HI-poor LTGs differs
statistically from the HI-rich ones.  To pursue this,  we measured  the
total extended flux  and the flux in the central 3 arcsec (mimicking the
SDSS spectral fibers) of HI-poor LTGs in the H$\alpha$ images.  The cumulative distribution of the ratio of the two
fluxes is given in Fig. \ref{fluxratio} (green histogram). This distribution
is significantly different from the one of HI-rich galaxies (blue histogram),
because the H$\alpha$ distribution of HI-poor galaxies is much more centrally concentrated. 
The HI-poor LTG galaxies appear to have some H$\alpha$ emission that is for the most part
confined to a central or even nuclear source\footnote{In order to protect against size effects,
we have recomputed the flux ratio in Fig. \ref{fluxratio} using a variable inner aperture of 3, 6 and 9
arcsec, each chosen according to the overall size of the galaxies. Our conclusions remain unchanged.}. 
A Kolmogorov-Smirnov test indicates that the probability that the two samples are drawn from the same parent population
is $2\times10^{-9}$\% , i.e., they differ significantly.
This is consistent with an  H$\alpha$ disk truncated by ram-pressure (Fumagalli \& Gavazzi 2008), such as in 
the prototype NGC 4569 in Virgo (Boselli et al. 2006) or in NGC 4848 in Coma (Fossati et al. 2012).
These cases are different from the HI-rich systems, which tend to
host H$\alpha$ emission extended across the entire disk (as discussed in Paper IV).   

\begin{figure} 
 \centering
\includegraphics[width=\columnwidth, trim = 0cm 0cm 0.5cm 0cm]{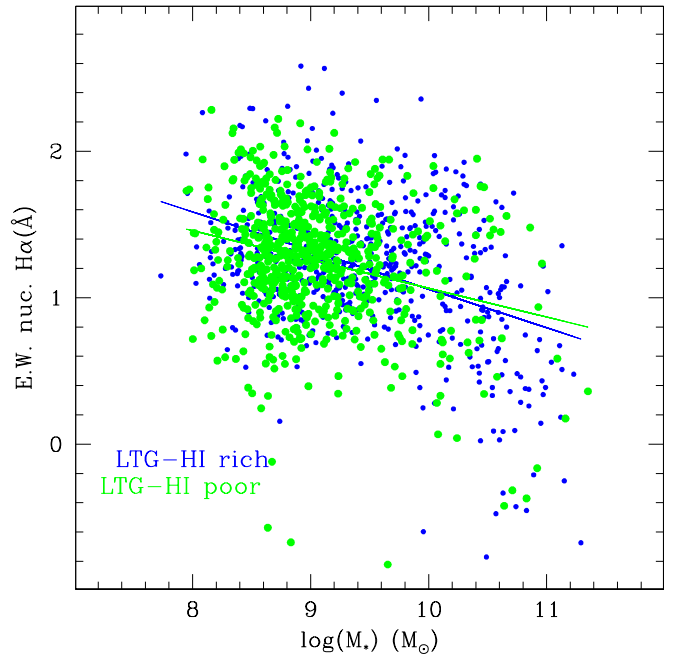}
\caption{
E.W. of the H$\alpha$ line as derived from nuclear SDSS spectra versus stellar Mass of LTG galaxies
(in subsamples 2). Galaxies are color-coded as follows:
 green=LTG-not in ALFALFA (HI-poor); blue=LTG-in ALFALFA (HI-rich). }
\label{HAnuc} 
\end{figure}

Figure \ref{HAnuc} shows that HI-poor and HI-rich LTGs have an identical  
intensity of the nuclear H$\alpha$ however. 
The figure shows the distribution of the
H$\alpha$ equivalent width (EW, a proxy for nuclear SSFR) as derived from the nuclear SDSS spectra versus the stellar
mass separately for HI-rich (blue symbols) and HI-poor (green symbols)
LTGs. Both subsamples show an identical mild decrease of the nuclear
H$\alpha$ EW  with increasing mass (similar to the decrease of the diffuse
SSFR with mass in Figure \ref{scale} d), but apart from that they have an
identical specific star formation activity in the nuclear region.
A Kolmogorov-Smirnov test indicates that the probability that the two samples are drawn from the same parent population
is 4.1 \%, i.e., they do not differ by more than 2 sigma.

In conclusion, HI normal, low-mass LTGs have extended star-forming disks. The star formation shrinks to the nucleus
when LTGs are affected by gas ablation, but on this scale the star formation occurs with the same intensity as in gas rich systems. 
Only ETGs become completely passive even in their nuclei (at stellar masses as low as $10^{9.5} ~M_{\odot}$ 
AGNs of high and low activity (LINER) are practically absent even among ETGs, 
see Gavazzi et al. 2011).
This provides solid evidence that the  mechanism that produces the quenching of the star formation proceeds outside-in,
as in the ram-pressure scenario. Other processes such as tidal interactions would truncate the star formation 
in the outer disks, but
would increase it as well on the nuclear scale (Kennicutt et al. 1987, see however Bretherton et al. 2010, who claimed some role of
tidal mechanisms in cluster galaxies). 
In contrast, milder ablation processes 
such as starvation could produce a cut-off of the cold gas supply an in turn a significant
quenching of the star formation even in galaxy groups (as supported by numerical 
simulations by Kawata \& Mulchaey  2008). However, this mechanism is not expected to 
reproduce the truncated HI disks (Cayatte et al. 1994, Chung et al. 2009)
and the truncated H$\alpha$ profiles (Koopmann \& Kenney 1998, 2004; Boselli et al. 2006)  observed in cluster galaxies.

\subsection{Local density or velocity dispersion?}

It is also interesting to check if the transformation just found, i.e., the progressive reduction of the star formation
with increasing density, 
occurs only in the environment of rich galaxy clusters or also in galaxy groups.
For this we studied a possible dependence of the fraction of HI-rich LTGs on the velocity dispersion $\sigma$
of the groups/cluster that host them.
This is because $\sigma$ traces the potential well of the hosting galaxy systems and, as a consequence, is
directly and physically linked to the strength of the ram-pressure process (Gunn \& Gott  1972, Domainko et al. 2006).
We developed a two-step automatic procedure that identifies and characterizes the galaxy groups in region 2. 
The first step identifies galaxies that can be considered tracers of galactic groups.
A tracer must lie in an overdensity with at least 20 members within a circle of 0.3 degree (530 kpc) radius. $\sigma$ 
and the mean recessional velocity $\langle Vel \rangle$ were estimated
for each tracer by averaging over its neighbors up to a maximum number of 30 galaxies. This allows for a more detailed description
of the ``local'' sigma value in the highest density regions like, e.g., the Coma cluster.
Once the tracers were identified, the code proceeded with step II, in which the membership 
of each individual LTG to 
a group/cluster was checked. To be hosted by a group, a galaxy ($i$) must lie within 
a projected separation of 0.5 deg (900 kpc)
and ($ii$) its velocity along the line of sight must not differ by more than 3 $\sigma$ from $\langle Vel \rangle$.
The code defines as ``field'' galaxy every object located more than 3 deg (5.3 Mpc) away from each overdensity.
This conservative constraint slightly reduces the number statistics, but prevents any contamination from 
galaxies in the outskirts of the overdensities themselves.

The procedures identified seven groups and the Coma cluster, as listed in Table \ref{groups}. 
The test discussed in Dressler \& Shectman (1988) confirmed the
substructures we found. This is to be expected because of the  velocity
criterion that we
enforced to define the group elements. This criterion selects
dynamically well-defined groups with a velocity dispersion quite a bit lower
than the Coma cluster itself, as shown in Figure \ref{sigma}.\\
\begin{table}[h!]
\caption{Properties of the seven groups found in region 2. }
\begin{center}
\begin{tabular}{lllllll}
\hline
Group	      & Name	& RA    & Dec & N & $<V>$ & $\sigma$ \\
\hline
gr1 & NGC 3651 &170.7608  & 24.210 &35 &    7745& 455\\
gr2 & NGC 4104 &181.6839  & 28.172 &27 &    8316& 672\\
gr3 & UGC 7115 &182.0195  & 25.260 &54 &    6627& 427\\
gr4 & NGC 4213 &183.8709  & 23.916 &12 &    6884& 287\\
gr5 & NGC 4555 &188.8925  & 26.573 &43 &    6615& 309\\
gr6 & UGC 8763 &207.7632  & 25.039 &37 &    8779& 230\\
gr7 & IC 4345  &208.8824  & 25.126 &23 &    8994& 260\\
\hline
\hline
\end{tabular}
\end{center}
\label{groups}
\end{table}
For each group we computed the fraction
of HI-rich LTGs with the associated Poissonian uncertainty.
The fraction of HI-rich LTGs for the entire late-type population
is plotted in Figure \ref{sigma}a
as a function of the velocity dispersion and in Figure \ref{sigma}b as a function of the local galaxy density
in three different environments: the large dot corresponds to the average fraction in the lowest 
density environment (field), the square is the fraction in the highest density 
environment (the Coma cluster) and the small dots correspond to seven individual groups, 
with the triangle showing their average value.
Apparently, the fraction of HI-rich LTGs decreases  with
increasing local density (Figure \ref{sigma}b) (approximately 1 $\sigma$ from isolated to groups and more than 4 $\sigma$
from groups to clusters).  
However, considering the large statistical uncertainty on the gas-rich fraction, 
  groups are truly intermediate between isolated and Coma
  galaxies, i.e., there is not definitive evidence that the entire decrease occurs in the densest environment.  
  
Conversely, when the fraction HI-rich
LTGs is plotted as a function of $\sigma$ (Figure \ref{sigma}a), it becomes
more clear that  their fraction drops at $\sigma>>100$ km s$^{-1}$, at the typical dispersion  of
Coma-like clusters 
(we caution, however, that the  analysis presented in Figure \ref{sigma} should be 
extended to comprise more Coma-like clusters
before any strong conclusion can be drawn).
Recent simulations by Bahe et al. (2012) have shown that the pressure exerted on the cold gas by the IGM is at 
least one order of magnitude higher in clusters than in groups. In addition, the efficiency of cold gas stripping
is relevant only within the virial radius even for a massive cluster like Coma (see however 
Merluzzi et al. 2013), which renders  a strong ram-pressure
efficiency in groups unlikely because it would require the infalling galaxy to pass very near the group center.

\section{Discussion and conclusion}
\label{discussion}

 The environmental conditions occurring in the vicinity of an evolved cluster of galaxies such as the 
 Coma cluster have well-known (Giovanelli \& Haynes 1985) catastrophic consequences on the HI content of their member galaxies. 
 However, HI observations deeper than ALFALFA are required to detect the pattern of HI deficiency (see Sections \ref{hidef} and \ref{undet}).
 At the distance of Coma ALFALFA is just sensitive enough  ($\log (M_{HI}/$M$_{\odot}) \geq 9$) to separate  
 optically selected LTGs into HI-poor and  HI-rich galaxies.
 
  The SSFR (see Figure \ref{colmagaa}c) of galaxies scales down with the galaxy stellar mass, 
  but for the dwarf regime ($8.5<\log(M_*/{\rm M_\odot})<9.5$) there is a clear tendency for the median $(g-i)_o$ color 
  to increase systematically (see Figure \ref{histo}a) along the ``morphological'' sequence constituted by \\
  1) HI-rich LTGs, the bluest objects with  median $(g-i_o)$=0.50 mag; \\
  2) HI-poor LTGs, with  median $(g-i_o)$=0.60 mag; \\
  3) PSBs, with median $(g-i_o)$=0.84 mag; \\
  4) ETGs, with median $(g-i_o)$=1.00 mag.\\ 
  The specific star formation rate decreases accordingly.
  Along this sequence there is a parallel sequence that refers to their nuclear properties:
  HI-rich LTGs have identical nuclear star formation activity as HI-poor LTGs (Figure \ref{HAnuc}), but
  their star formation extends over the whole disks, while
  HI-poor LTGs harbor some star formation exclusively on a circumnuclear scale (Figure \ref{fluxratio}). The star formation 
  becomes entirely suppressed, even on nuclear scale in PSBs and ETGs.
  Whatever mechanism  quenches the star formation, it proceeds outside-in, as in the ram-pressure scenario
  (see also Boselli et al. 2006, 2009).
   \begin{figure*}
\centering
\includegraphics[width=\columnwidth, trim = 0.5cm 0.5cm 1cm 0cm]{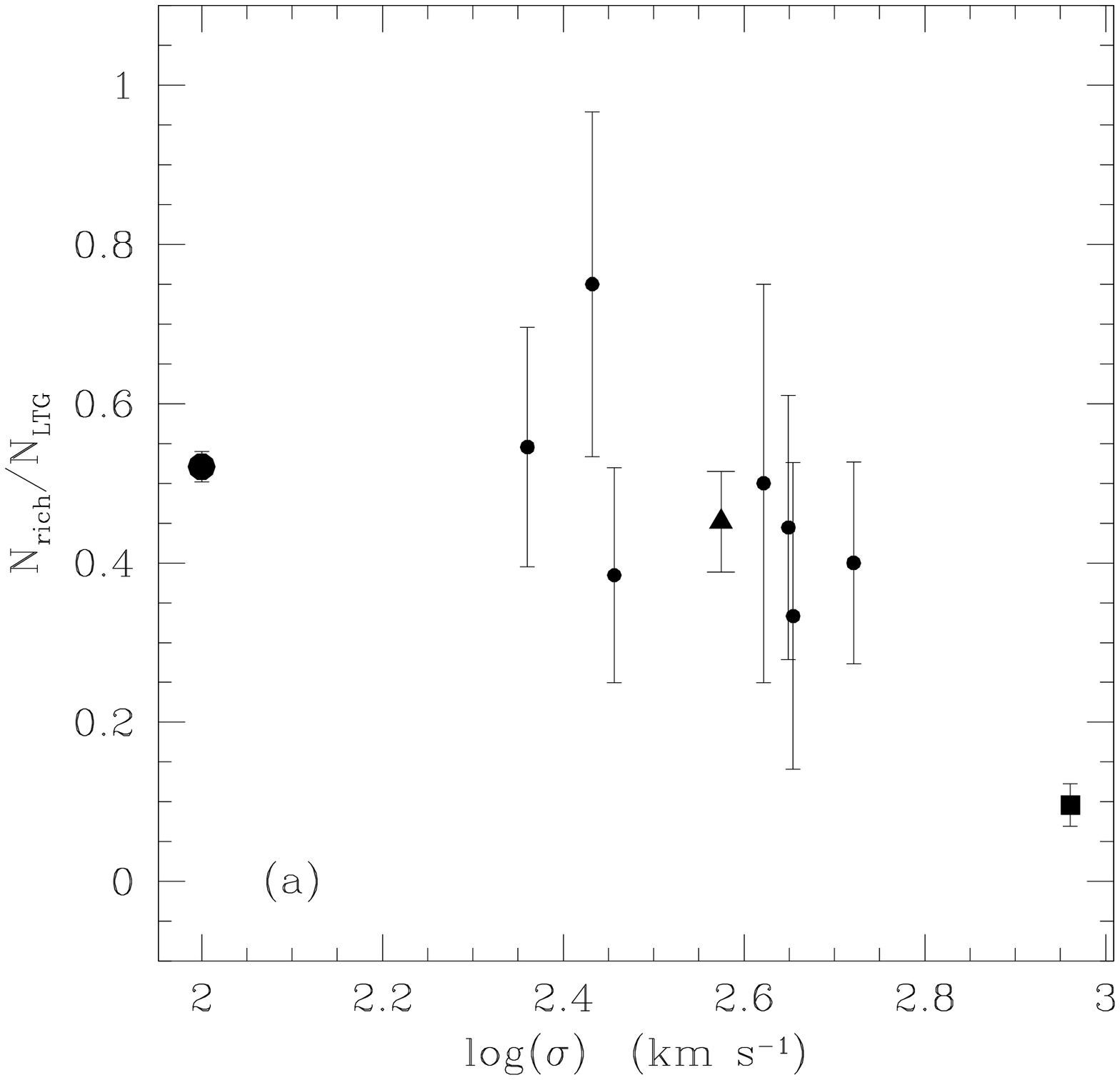}
\includegraphics[width=\columnwidth, trim = 0.5cm 0.5cm 1cm 0cm]{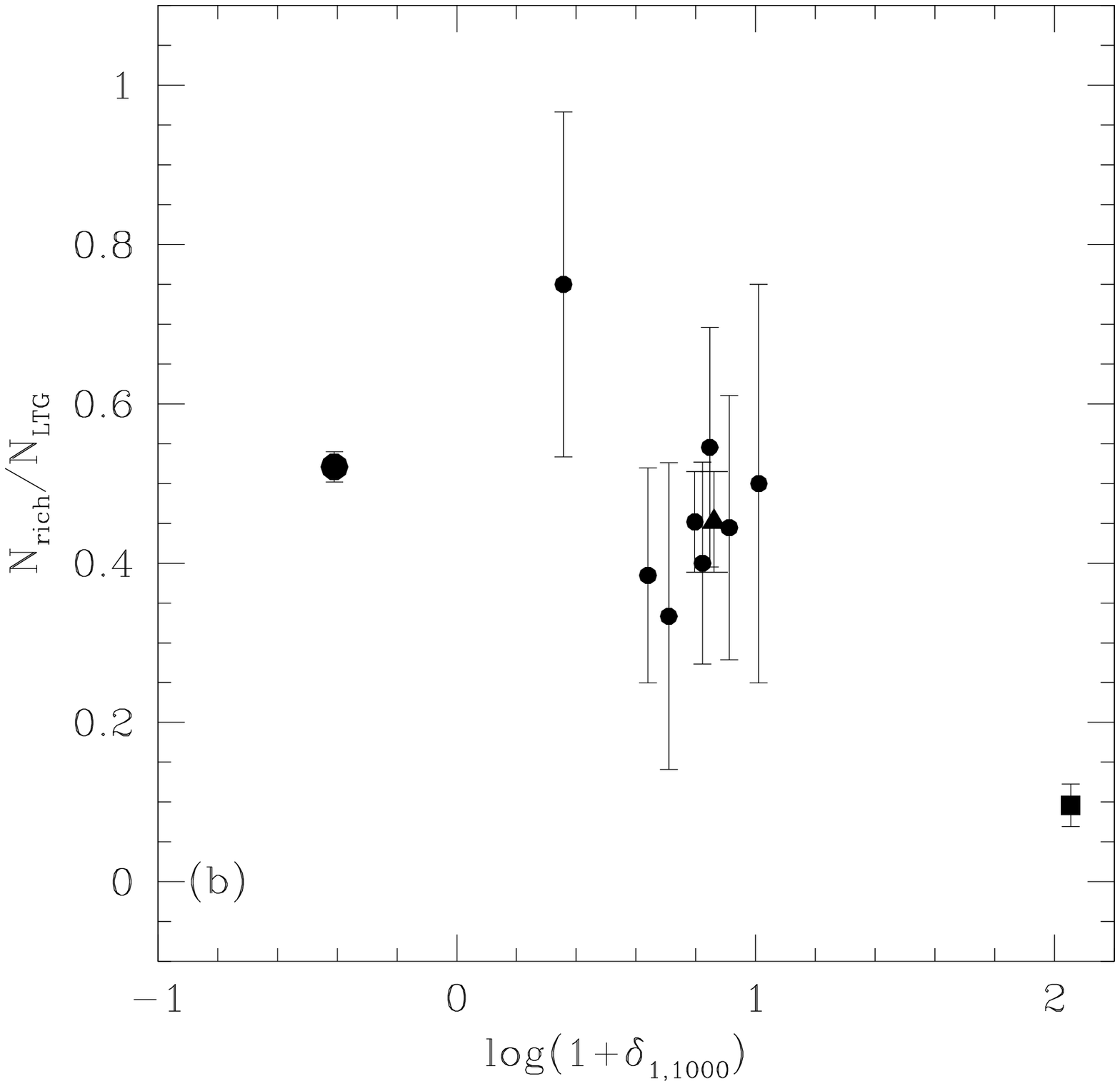}
\caption{
The number ratio of HI-rich LTGs over all LTGs as a function of the velocity dispersion (panel a) and of the local density (panel b) of 7 groups
(small dots) and their average (triangle), the Coma cluster (square) and the field (large dot). The last point has been plotted
at 100 $\sigma= \rm km ~s^{-1}$ to represent that, even in the field, galaxies are subject to random motions.
}
\label{sigma} 
\end{figure*} 

  A similar $(g-i)_o$ - stellar mass sequence is obtained by dividing galaxies into
  bins of increasing local density (Figure \ref{colmagaa}b 
  and \ref{histo}b).  It appears indisputable that the local density
    has a main impact on the gas content and SF of galaxies, contributing to their morphological migration,
  i.e., HI-rich LTGs represent about 60\% of the entire LTG population at low
  and intermediate galaxy density (comprising the cosmic filaments up to the groups)
  and drop to zero in the core of the Coma cluster. Conversely, HI-poor LTGs
  (representing the other 40\% of the LTGs in isolation) increase in frequency
  with increasing density, becoming the majority of all LTGs in the core of
  Coma.  This agrees with the analysis of Fabello et al. (2012) based on a
  $\log (M_{*}/$M$_{\odot}) \geq 10.0$ sample.
  
    It is very tempting to interpret all these features as the
  results of a fast environmentally driven process that superposes to the
  standard and long-term evolutionary track that acts on field and group
  galaxies.  Galaxies infalling toward a cluster are evolved by a process that
  causes some gas ablation, proceeding outside-in in two steps: 1) first a
  significant quenching of the global star formation is produced in the disk,
  resulting in redder colors, but without affecting the circumnuclear star formation; 2) the
  quenching of SF activity proceeds to the nuclear scale (transforming star-forming 
  nuclei into passive ones, passing through the PSB phase). On a
  similar time scale, dwarf galaxies are transformed from LTGs to ETGs.
 The time scale for gas depletion is short enough ($\sim$ 100-300 Myr) to quickly
 transform HI-rich LTGs into anemic LTGs, which end up as gas-free, passive
 dEs. This is precisely the mechanism invoked by Boselli et al. (2008) and by
 Gavazzi et al. (2010) to migrate dwarf star-forming systems into dEs in the
 neighborhood of rich galaxy clusters (see also Cortese \& Hughes 2009 and Cortese et al. 2011).  
 Meanwhile the ram pressure mechanism, owing to its short time scale, is also 
 capable of producing a k+a signature in their spectra, thus 
 enhancing the number of galaxies presenting some PSB phase toward rich
 clusters (Gavazzi et al. 2010; Wong et al. 2012) and making the number of
 active nuclei (both star forming and AGNs of various forms) decrease near
 the center of rich clusters (Gavazzi et al. 2011). 

The dependence of this quenching process on the galaxy density and in
particular on the galaxy velocity dispersion (it acts selectively on galaxies
hosted in galaxy clusters, see Figure \ref{sigma}), together with the evidence that it proceeds from outside-in, 
make the ram-pressure mechanism (Gunn \& Gott 1972) our favorite
interpretation over strangulation (Larson et al. 1980) 
This conclusion should be taken with the grain of salt because it is  based
on one cluster (Coma) and should be confirmed with similar analysis of more clusters.
In support of our conclusion, Merluzzi et al. (2013) found evidence for ram-pressure stripping 
well beyond 1 Mpc from the center of Abell 3558. In contrast the numerical simulations of
Kawata \& Mulchaey (2008) support the starvation mechanism as capable of cutting off the gas supply, hence 
of quenching the star formation even in galaxy groups.

We finally note that this fast evolutionary scenario could be used
to constrain the build-up of the Coma cluster with cosmic time at the expenses of infalling galaxies, following 
the argument of Adami et al. (2005) for Coma and of Boselli et al. (2008) for Virgo.
Assuming naively that the infall rate was constant over the past few Gyrs,
this quantity can be estimated from the number of galaxies in each specific morphological stage divided by the typical time scale of 
each stage (to be precise, since this estimate is very sensitive to the sample completeness, the argument
provides lower limits to the infall rate).
Restricting this to galaxies with $\log(M_*/$M$_{\odot})> 9$, to be consistent with a similar calculation 
carried out in Paper II for Virgo,
we classified 26 HI-rich LTGs within the cluster. Assuming a characteristic time scale for the
complete stripping of gas due to ram pressure of  300 Myr for massive galaxies such as 
NGC 4569 (Boselli et al. 2008)\footnote{This estimate is reduced to 100 Myr for dwarf objects such as VCC 1217 
(Fumagalli et al. 2011) or VCC 1249 in the M49 group (Arrigoni Battaia et al. 2012).}, we obtain a flux of infalling
LTGs of $F_{\rm gal}=86$ Gyr$^{-1}$.
We can cross-check this result with the independent constraint derived from
the number of galaxies in the HI-poor phase (70) 
(that still show some circumnuclear star formation probably sustained by residual 
 nuclear HI and $\rm H_2$ anchored to the deep central potential well), 
divided by the time-scale of $\rm H_2$ ablation. 
Assuming that a complete HI and $\rm H_2$ removal will occur due to ram pressure during the second passage through the cluster center
on a time-scale of $0.7 \pm 0.4$ Gyr (see Paper II), the infall rate is
$F_{\rm gal}=100$ Gyr$^{-1}$. 
In passing we note that the time scale estimate we used is considerably shorter than that for complete consumption of the $\rm H_2$  
 due to star formation of $\sim 2\pm 1$ Gyr (Bigiel et al. 2008). 

A similar argument holds for the 32 PSB galaxies observed in Coma. 
The PSB phase is expected to last 
up to 1.5 Gyr (Poggianti et al. 2009), but around
500 Myr in a blue continuum regime (see Fig. 20 in Boselli et al. 2008), such as in the 32 PSB galaxies analyzed here. 
Again, the galaxy flux toward Coma would be $F_{\rm gal}=64$ Gyr$^{-1}$, 
consistent with the previous estimates. Assuming that
the age of the Coma cluster is about 7.5 Gyr (its formation took place approximately at $z$=1, 
Wechsler et al 2002), and that it is composed of
about 750 galaxies, the infall rate would become $F_{\rm gal}=130$ Gyr$^{-1}$.
All ''order of magnitude'' estimates are consistent with one another within
30 percent from $F_{\rm gal}=100$ Gyr$^{-1}$, i.e.,
a constant galaxy flow building the cluster over a large portion of the cosmic time, as was concluded by Adami et al. (2005).
Consistent infall rate, sustained for 2 Gyr, has been claimed to feed the younger Virgo cluster (Paper II).

The infall rate estimate can be checked for consistency with semi-analytic models of halo formation,
e.g., Wechsler et al (2002). 
From their Figure 3 we derive that
cluster-like halos accreted 70\% of their final mass in the last 7.5 Gyr 
(the assumed age of the Coma cluster).
If one takes for Coma a current halo mass of approximately $10^{14}~\rm M_\odot$,
the average growth is  $10^{13} ~\rm M_\odot~Gyr^{-1}$. 
Considering (very crudely) that this accretion is contributed by galaxy halos of 
$10^{11}~\rm M_\odot$ each, the derived accretion rate is 100  
galaxies per Gyr on average, in agreement with our previous estimate.  
    \begin{table}[h!]
    \caption{Build-up of the Coma cluster from various estimators.}
    \begin{center}
    \begin{tabular}{lllll}
    \hline
         &All   		     	& PSB			& HI-poor	     & HI-rich  	               \\
    \hline
    N ($\log(M_*/$M$_{\odot})> 9$)      & 750		&  32			& 70		    & 26		\\
    t (Gyr)		                & 7.5 $\pm$ 1	&  0.5 $\pm$ 0.2	& 0.7 $\pm$ 0.4     & 0.3 $\pm$ 0.1	\\
    Infall rate (Gyr$^{-1}$)            & 100 $\pm$ 14  &  64 $\pm$ 28	        & 100 $\pm$ 58	    & 86 $\pm$ 33 	    \\
    \hline
    \hline
    \end{tabular}
    \end{center}
    \label{fit}
    \end{table}

    \appendix
    \section{empirical correction for internal extinction}
    \label{correction}
   
\begin{figure} 
 \centering
\includegraphics[width=\columnwidth,trim = 0.5cm 1.2cm 8cm 0cm]{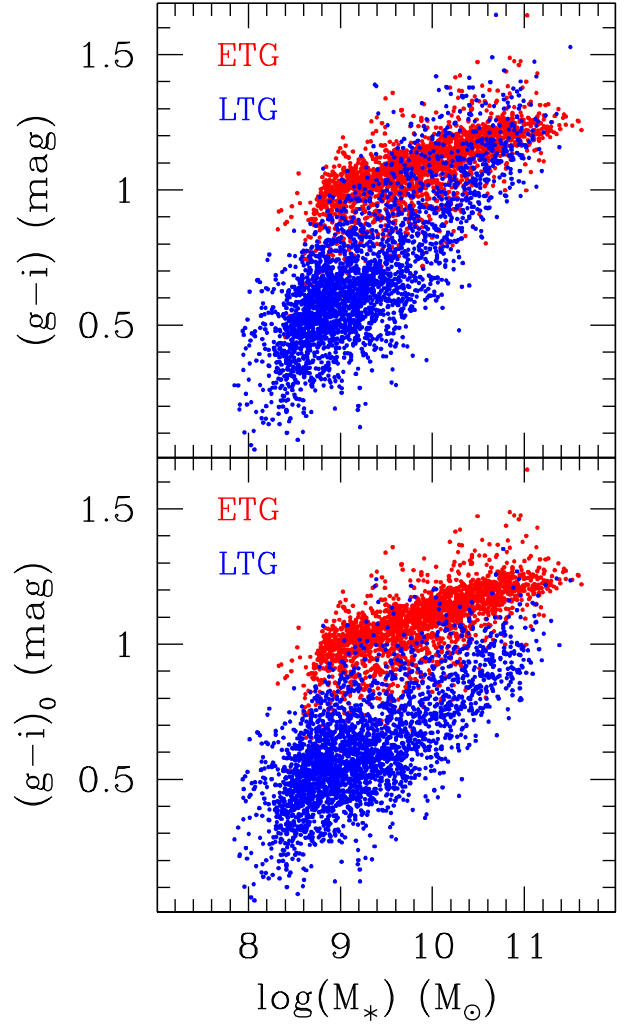}
\caption{
(top): observed color ($g-i$) -stellar mass diagram of all 5026 optically selected galaxies from SDSS 
(subsamples 1+2) divided
by morphological types: ETGs (dE-S0a, red), LTG (Sa-Im-BCD, blue). Many high-mass LTGs overlap with, or are even redder
than, ETGs of similar mass. 
 (bottom): same as above but $(g-i)_o$ is corrected for inclination as described in this appendix. The correction 
 is visible at high mass where LTGs have now bluer colors than ETGs. The blue and red sequence become more
 parallel and a significant green valley opens up between them.
}
\label{colmag}
\end{figure}

\begin{figure} 
 \centering
\includegraphics[width=\columnwidth, trim = 0cm 0.5cm 0cm 0cm]{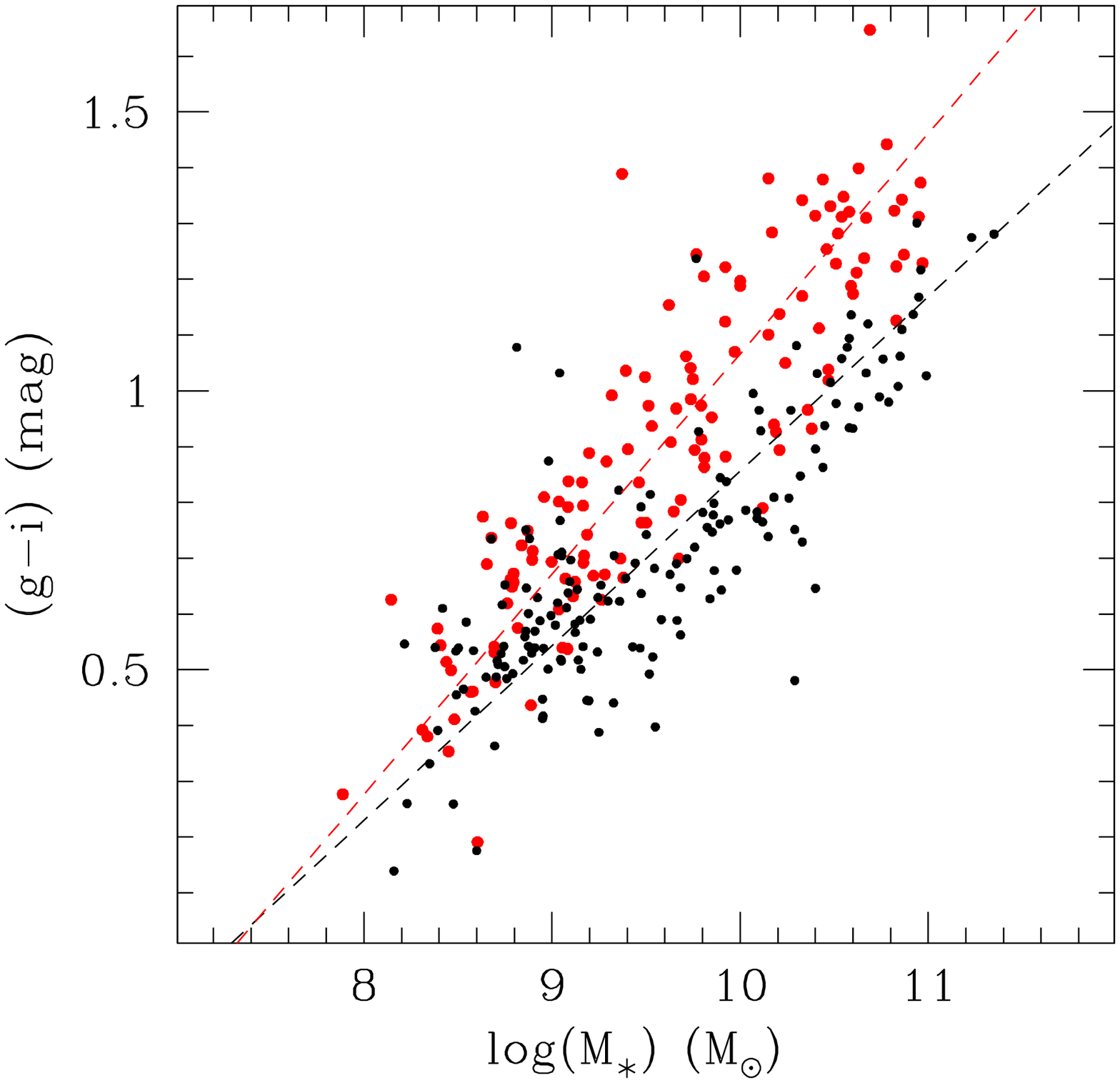}
\includegraphics[width=\columnwidth, trim = 0cm 0.5cm 0cm 0cm]{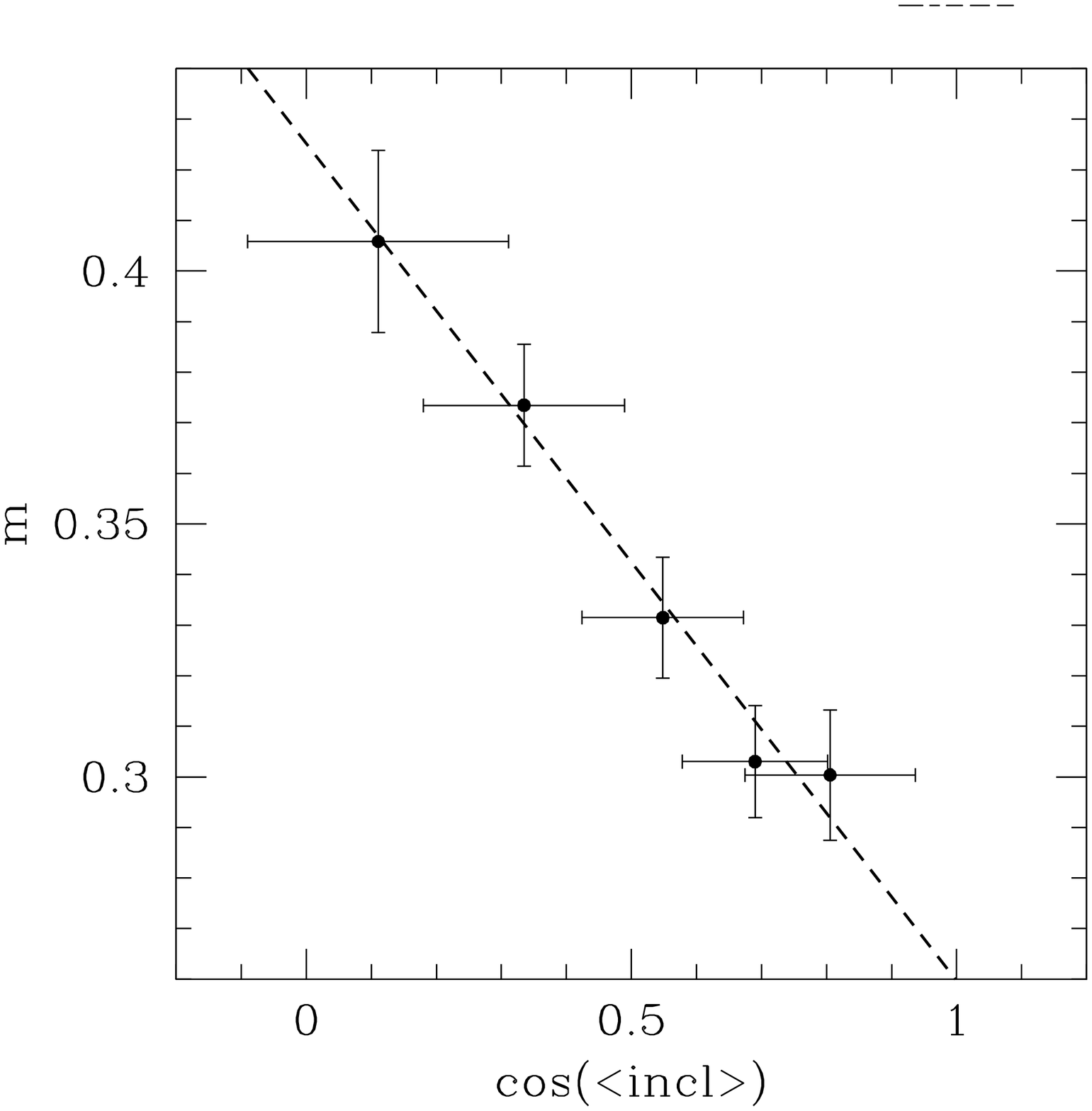}
\caption{
(Top panel) Observed color ($g-i$) -stellar mass diagram (subsamples 1+2) of edge-on LTGs ($incl>80$, red) 
and of face-on LTGs ($incl<30$, black), each with the linear best fit (dashed line).
(Bottom panel) The slope of the fit to the color ($g-i$) -stellar Mass relation of disk galaxies (Sa-Sdm) 
(see Figure \ref{colincl}) as a function of galaxy inclination.}
\label{colincl}
\end{figure}

An exhaustive treatment of the internal extinction correction to the $u,g,r,i,z$ 
magnitudes is beyond the scope
of this work and will be addressed in a forthcoming paper. 
We concentrate on a less ambitious task, i.e., on the empirical method 
for correcting the color magnitude relation for the effects of internal extinction.
The color ($g-i$) versus stellar mass diagram (see Figure \ref{colmag}, color-coded according to the Hubble type) 
composed of all optically selected galaxies (regions 1 and 2) 
contains a well-developed red-sequence (e.g. Hogg et al. 2004) mostly composed of ETGs 
along with the so-called blue cloud, mostly composed of LTGs.
When magnitudes are not corrected for internal extinction the two color sequences, 
that are well separated at low mass by the green valley, overlap one another at the 
high-mass end. This is an obvious consequence of the increasing internal extinction suffered by disk galaxies
of increasing mass.
Given the abundant number statistics offered by SDSS, it is possible to investigate this effect as a function of  
galaxy inclination in the plane of the sky.
For disk galaxies  (Sa through Sdm)
the inclination $incl$ with respect to the line of sight was computed following Solanes et al. (1996)
assuming that spirals are oblate spheroids of intrinsic axial ratio $q$:
\begin{equation}
\cos^2(incl) = \frac{(b/a)^2 - q^2}{1 - q^2} ,
\end{equation}
where $a$ and $b$ are the major and minor axes. Intrinsic axial ratios were assumed to be equal
to the modal values of the distribution of true ellipticities of galaxies of different Hubble types
in the Second Reference Catalog of Bright Galaxies (de Vaucouleurs, de Vaucouleurs, \& Corwin 1976). 
For the giant spirals these are
$q=0.32$ for Sa, $q=0.23$ for Sab, and $q=0.18$ for Sb-Sc. Whenever $b/a<q$, $incl$ is set to $90^o$.
Irregulars+BCD (mostly low-mass objects) do not have this parameter defined, because their axial ratio does not 
give a measure of the inclination of the disk, but of their intrinsic triaxiality.\\
Figure \ref{colincl} (top panel) shows the $g-i$ color versus stellar mass diagram for LTGs in regions 1+2 considering only 
nearly edge-on ($incl>80$) and nearly face-on ($incl<30$) galaxies, with the best linear fit. 
It is evident that the fits cross each other near log$M_*=8.0$, i.e., the extinction correction is null near log$M_*=8.0$,
and it increases with increasing stellar mass as a function of the inclination.
The slope of the relation
as a function of the cosine of the inclination is given in Figure \ref{colincl} (bottom panel).
This relation can be used to extrapolate $g-i$ to the face-on value, or to obtain the color correction 
for internal extinction:
\begin{equation}
(g-i)_o = (g-i) - \Bigl\{ +0.17 \cdot [1-\cos(incl)] 
\cdot \Bigl[\log\Bigl(\frac{M_*}{M_\odot}\Bigr)-8.19\Bigr]\Bigr\}, \
\end{equation}
where $incl$ is the galaxy inclination.  
After applying the empirical internal extinction correction 
the resulting slope of the LTG sequence becomes similar to the one of ETGs, and the green-valley opens up
over the full interval of masses (Figure \ref{colmag} - bottom panel). Note the difference between the correction method proposed here 
(which depends on a combination of $M_*$ and $incl$) and the one proposed by Yip et al. (2011), which 
purely depends on $incl$.

\section {Scaling relations}
\label{scale}

 H$\alpha3$, which is based on an ALFALFA-selected sample in the Coma supercluster 
 provides us with insufficient coverage of HI mass, stellar mass, and SFR because of
 the shallowness of ALFALFA at the distance of Coma.
 Therefore the analysis of the scaling relation among the HI gas content and the stellar mass of galaxies 
 cannot be undertaken with a sufficient dynamical range of HI mass, stellar mass, and SFR.
 However by adding the results of the present survey with similar ones obtained in Paper II for the Local Supercluster
 (where the ALFALFA sensitivity is 35 times better), we broaden the dynamic range to four decades in stellar mass,
 obtaining the relations shown  in Figure \ref{scale}, whose 
 linear regression parameters are listed in Table \ref{fit}.
 The data obtained in this paper for the Coma supercluster  for galaxies with $Def_{HI} \leq 0.2$ are plotted
 separately (blue symbols) from those in the Local Supercluster with $Def_{HI} \leq 0.3$ taken from paper II (light gray).  
 AGNs of various levels of activity  are identified in Figure \ref{scale} (with asterisks) according 
 to the criteria of Gavazzi et al. (2011).
  
 We begin by studying the relation between $M_{HI}$ and $M_*$, as shown in
 Figure \ref{scale} (a).  The diagonal lines represent the sensitivity limit of
 ALFALFA, computed for galaxies with an inclination of 45 deg (see Section
 \ref{radioselection}). The dotted line is computed for the distance of the Virgo
 cluster (17 Mpc), while the dashed one is for the distance of Coma (100 Mpc).
 Clearly the observed distribution for Coma lies very close to
 the line of limiting sensitivity, making the observed slope (0.34) spuriously
 shallow because galaxies near  $\log (M_*/$M$_{\odot})\sim 8$ and $\log (M_{HI}/$M$_{\odot})\sim 8$,
 which are present locally,  would not be detected at the distance of Coma.
 However, at the high-mass end of the stellar mass function, where the sampling 
 is more abundant in Coma than locally because of the larger volume,
 no selection effects would prevent us from detecting galaxies if they 
 had the same steep slope as extrapolated from the Local Supercluster.
 Therefore we conclude that the change of slope seen in  Figure \ref{scale} (a) near
 $\log (M_*/$M$_{\odot})\sim 9$ is real.
 A relation similar to the one in  Figure \ref{scale}(a) and (b), i.e., significant flattening 
 above $\log( M_*/$M$_{\odot})\sim 8$, has been obtained by Huang et al. (2012) using the whole
 $\alpha$.40 sample. They interpreted the flattening as a possible evidence of increasing gas loss due to AGN feedback
 with increasing mass. 
 
 We note that the observed relation, both in the steep part (0.47) and in the shallow (0.32),
 is nonetheless significantly shallower
 than the direct proportionality, as already discussed in Gavazzi et
 al. (2008) and in Paper II, reflecting a genuine non-proportionality between
 the two quantities.  When the HI gas fraction $M_{HI}$ /$M_*$ is plotted as a
 function of the stellar mass (Figure \ref{scale} b), the fraction of gas
 decreases by approximately 2 orders of magnitude with increasing galaxy mass,
 from $\log( M_*/$M$_{\odot})\sim 7.5$ to $\sim 11.5$. In other
 words, $10^{7.5}$ M$_\odot$ galaxies contain ten times more gas than
 stars. This ratio is reduced to less than 10\% in $10^{11.5}$ M$_\odot$ systems.
 This basic result agrees perfectly with what was found in the
 (optically selected) GASS survey by Catinella et al. (2010) in the stellar
 mass range $10^{10}<M_*<10^{11.5}$ M$_\odot$ and by Cortese et al. (2011)
 for HI-normal galaxies in the Herschel Reference Survey (HRS; Boselli et
 al. 2010).  It is also consistent with the $downsizing$ scenario (Gavazzi
 1993, Gavazzi et al. 1996, Cowie et al. 1996; Gavazzi \& Scodeggio 1996, Boselli et al. 2001,
 Fontanot et al. 2009), where progressively more massive galaxies have less
 gas-to-star ratio at $z$=0 because they have transformed most of their gas into stars
 at higher $z$, while dwarf LTGs retain large quantities of hydrogen capable
 of sustaining the star formation at some significant rate at the present
 cosmological epoch.  
  
 Similarly, we find that the relation between the SFR and the stellar mass (Figure \ref{scale} c) 
 and that of the specific SFR with stellar mass (d) are consistent with a change of slope
 around $\log( M_*/$M$_{\odot})\sim 9$. Below that threshold a steeper dependence (0.8, i.e.,
 almost of direct proportionality) is
 traced by the local galaxies. Above this limit the relation is much shallower (0.33) 
 and is followed by the more distant Coma galaxies. 
 Consequently, in the range of stellar mass covered by the Coma supercluster
 the specific star formation rate decreases significantly with increasing mass, as found by 
 Boselli et al. (2001),  Bothwell et al. (2009) and by Schiminovich et al. (2010). This is again
 consistent with the "downsizing" scenario.
 At lower mass the relation flattens out and the scatter increases by 2 orders of magnitude, as noted by other authors 
 (Lee at al. 2007, Bothwell et al. 2009).
 The curvature in Figure \ref{scale} resulting from the superposition of the local and Coma superclusters 
 is consistent with the change of slope found by Huang et al. (2012) in  $\alpha$.40.
\begin{figure*}
 \centering
 \includegraphics[width=\columnwidth]{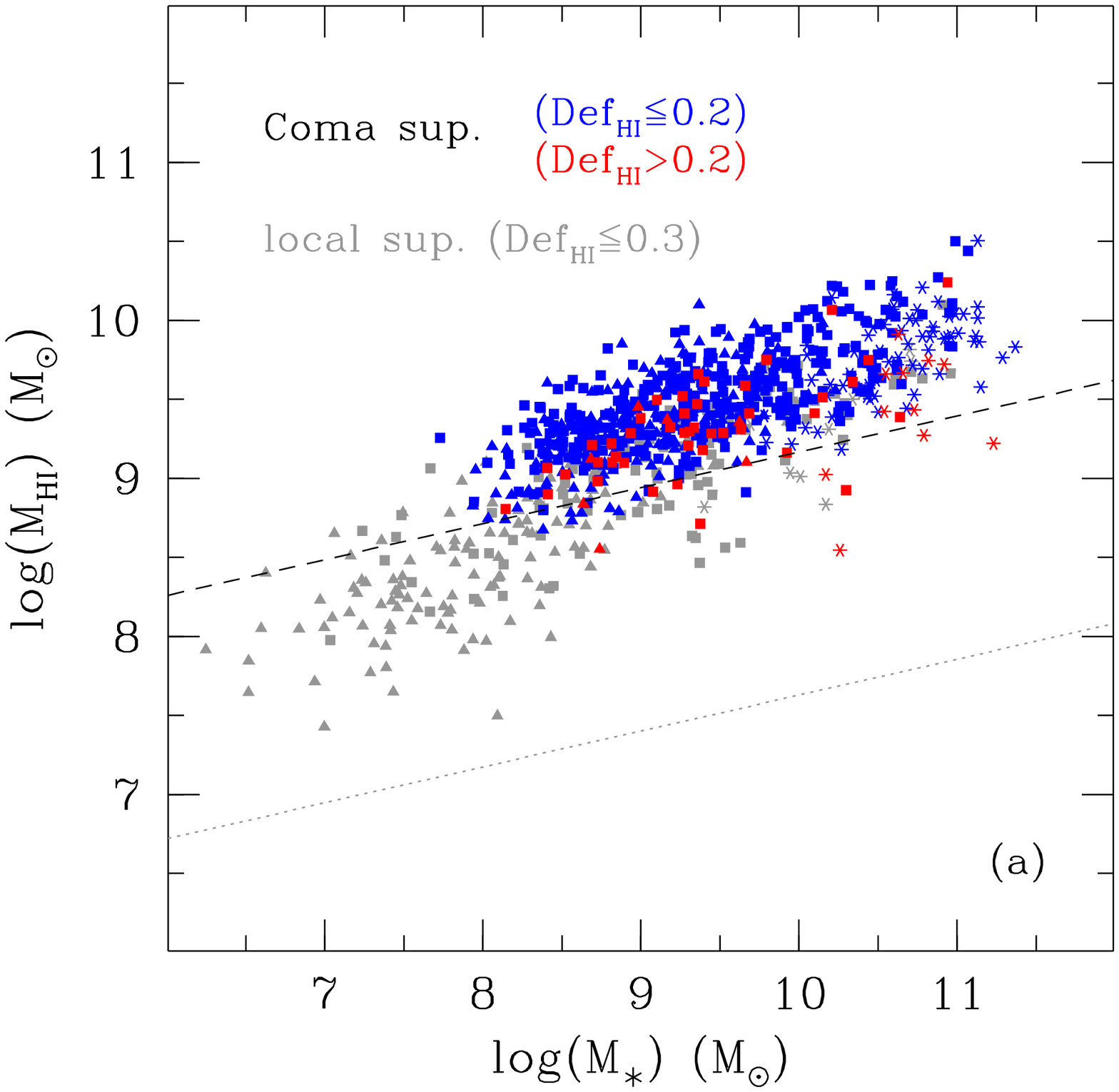}
  \includegraphics[width=\columnwidth]{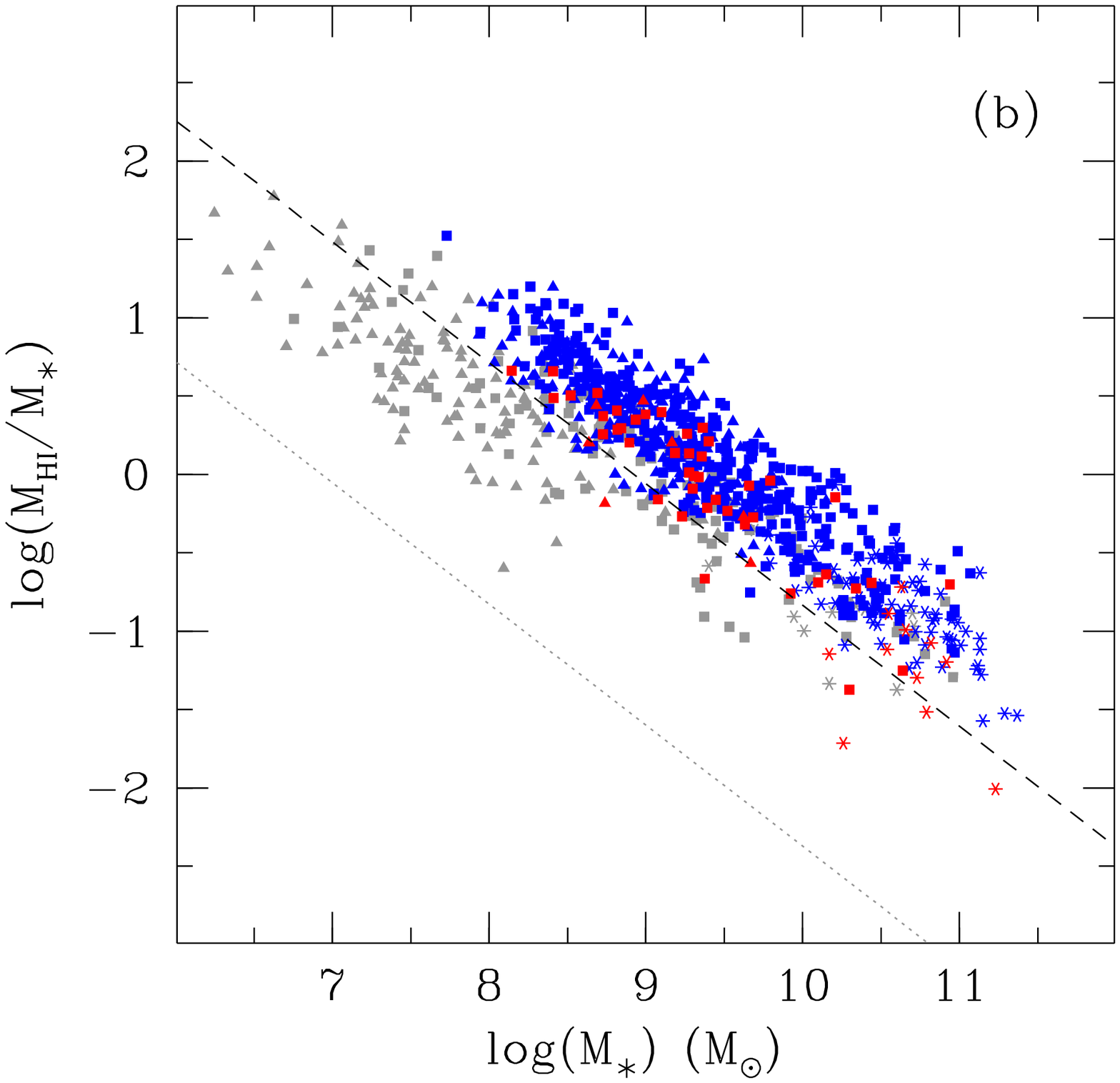}
   \includegraphics[width=\columnwidth]{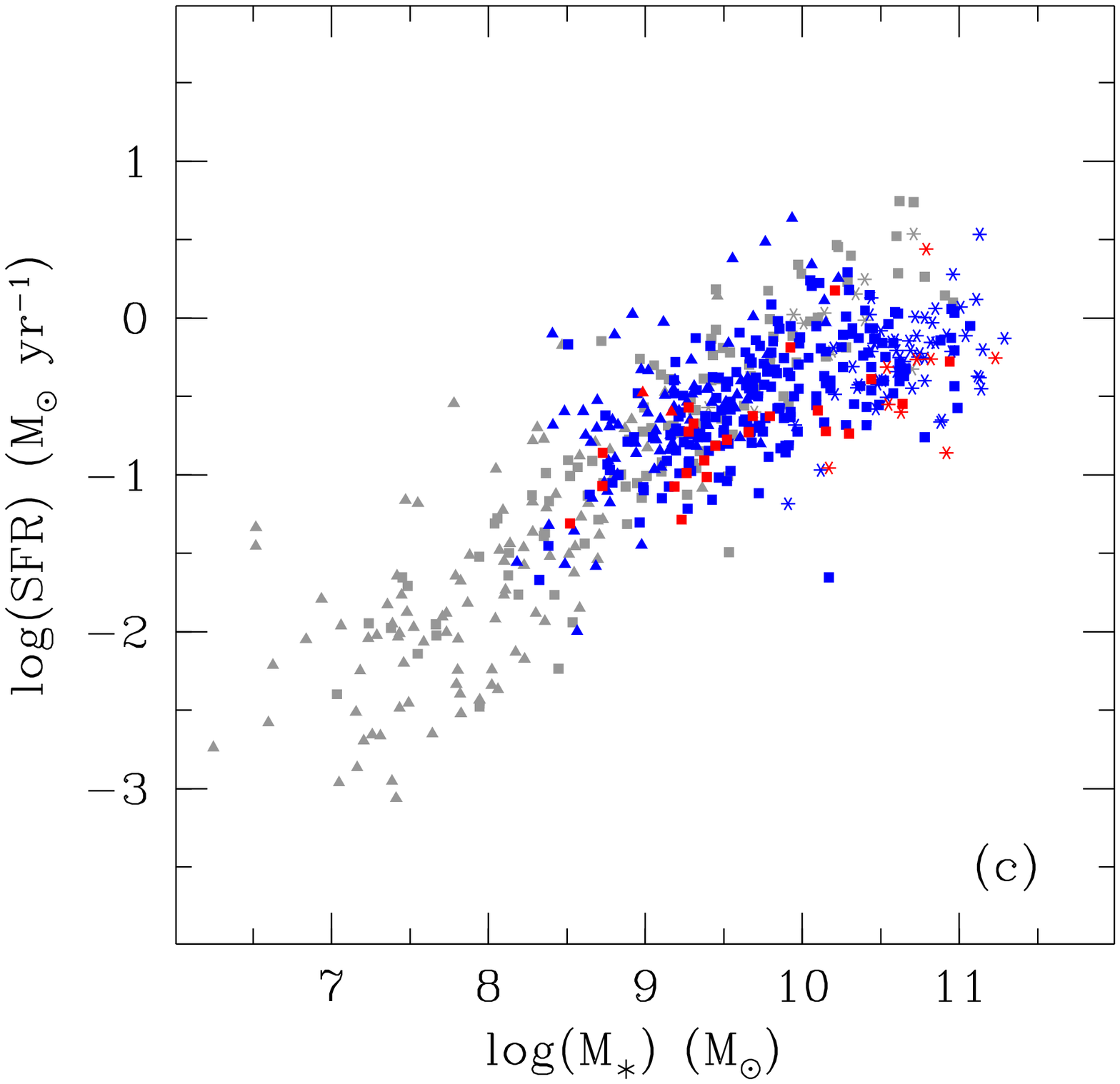}
    \includegraphics[width=\columnwidth]{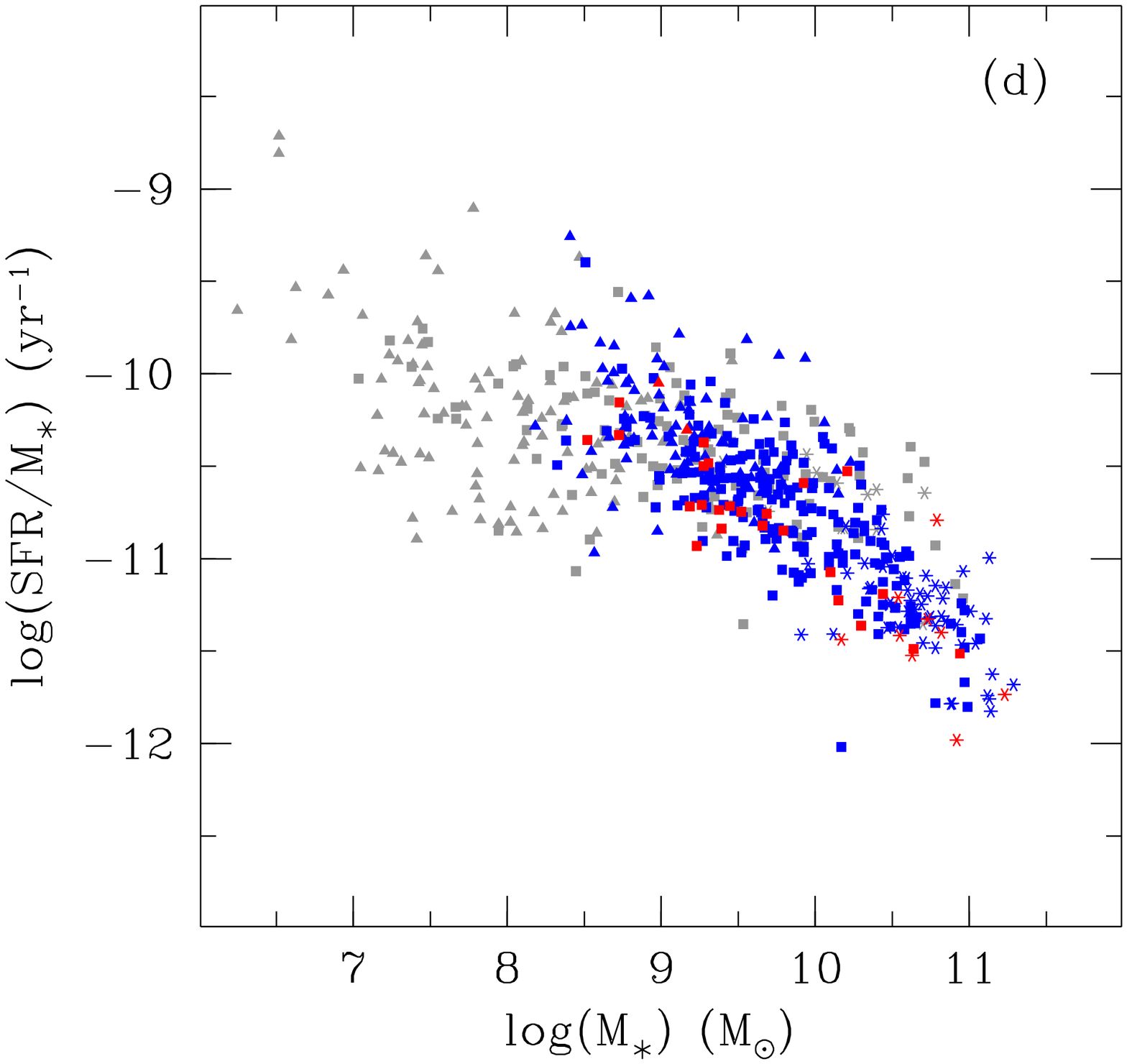}
  \caption{Principal scaling relations between $M_{HI}$, $M_*$, and SFR 
   in  the Coma supercluster covered by ALFALFA.
  Blue: $Def_{HI} \leq 0.2$;  red: $Def_{HI} \geq 0.2$  compared to the $Def_{HI} \leq 0.3$ in the Local Supercluster
  (grey, Paper II). Symbols are assigned according to the
  morphology: spirals (Sa-Sdm): squares; Irr-BCD: triangles; AGN: asterisks. 
  Top-left panel: the $M_{HI}$ versus $M_*$ relation
  The diagonal lines represent the limiting sensitivity of ALFALFA (computed for a mean $incl=45^o$) 
  at the distance of Virgo (gray dotted)
  and of Coma (black dashed). The shallow sensitivity of ALFALFA limits significantly the detection of galaxies at 
  the distance of Coma, in particular with decreasing stellar mass. 
  Top-right panel: the HI  gas fraction ($M_{HI}$/$M_*$) versus $M_*$. 
  Bottom-left panel: the relation between the SFR and the stellar mass.
  Bottom-right panel: the specific star formation rate versus stellar mass.}
 \label{scale}
 \end{figure*}

 \begin{table}[h!]
\caption{Linear regression coefficients of the relations in Figures 
\ref{scale}(a) through (d) (only for $Def_{HI}<0.2$)
obtained using the bisector method (mean coefficients 
of the direct and the inverse relation). C=Coma, V=Virgo (Paper II)
}
\begin{center}
\begin{tabular}{lrlr}
\hline
Linear regression                           & $r$          & sample     & Fig.\ref{scale}  \\
\hline
  log$M_{HI}=0.34\times$ log$M_*+6.25$      & 0.76           & C	    & a\\
  log$M_{HI}=0.47\times$ log$M_*+4.71$      & 0.85           & V	    & a\\
  log$M_{HI}=0.52\times$ log$M_*+4.54$      & 0.84           & V+C          & a\\
  log$M_{HI}/M_*=-0.68\times$ log$M_*+6.50$ & -0.92          & C	    & b\\  
  log$M_{HI}/M_*=-0.69\times$ log$M_*+6.12$ & -0.90          & V	    & b\\  
  log$M_{HI}/M_*=-0.71\times$ log$M_*+6.58$ & -0.86          & V+C          & b\\
  log$SFR=0.33\times$ log$M_*-3.71$         & 0.52           & C	    & c\\
  log$SFR=0.80\times$ log$M_*-8.04$         & 0.88           & V	    & c\\
  log$SFR=0.64\times$ log$M_*-6.75$         & 0.85           & V+C          & c\\
  log$SSFR=-0.70\times$ log$M_*-3.43$       & -0.79          & C	    & d\\
  log$SSFR=-0.26\times$ log$M_*-7.53$       & -0.53            & V	    & d\\
  log$SSFR=-0.41\times$ log$M_*-6.23$       & -0.72          & V+C          & d\\
\hline
\hline
\end{tabular}
\end{center}
\label{fit}
\end{table}

\begin{acknowledgements}
The authors would like to acknowledge the work of the entire ALFALFA collaboration team
in observing, flagging, and extracting the catalog of galaxies used in this work
and thank Shan Huang for providing original data.
This research has made use of the GOLDmine database (Gavazzi et al. 2003)
and of the NASA/IPAC Extragalactic Database (NED) which is operated 
by the Jet Propulsion Laboratory, California Institute of Technology, under contract with the
National Aeronautics and Space Administration. 
We wish to thank the unknown referee whose criticism helped us to improve the manuscript.
Funding for the Sloan Digital Sky Survey (SDSS) and SDSS-II has been provided by the 
Alfred P. Sloan Foundation, the Participating Institutions, the National Science Foundation, 
the U.S. Department of Energy, the National Aeronautics and Space Administration, 
the Japanese Monbukagakusho, and 
the Max Planck Society, and the Higher Education Funding Council for England. 
The SDSS Web site is \emph{http://www.sdss.org/}.
The SDSS is managed by the Astrophysical Research Consortium (ARC) for the Participating Institutions. 
The Participating Institutions are the American Museum of Natural History, Astrophysical Institute Potsdam, 
University of Basel, University of Cambridge, Case Western Reserve University, The University of Chicago, 
Drexel University, Fermilab, the Institute for Advanced Study, the Japan Participation Group, 
The Johns Hopkins University, the Joint Institute for Nuclear Astrophysics, the Kavli Institute for 
Particle Astrophysics and Cosmology, the Korean Scientist Group, the Chinese Academy of Sciences (LAMOST), 
Los Alamos National Laboratory, the Max-Planck-Institute for Astronomy (MPIA), the Max-Planck-Institute 
for Astrophysics (MPA), New Mexico State University, Ohio State University, University of Pittsburgh, 
University of Portsmouth, Princeton University, the United States Naval Observatory, and the University 
of Washington.\\
G. G. acknowledges financial support from the Italian MIUR PRIN contract 200854ECE5.
R.G. and M.P.H. are supported by US NSF grants AST-0607007 and AST-1107390
and by a Brinson Foundation grant.
Support for M.F.  was provided by NASA through Hubble Fellowship grant HF-51305.01-A awarded by the 
Space Telescope Science Institute, which is operated by the Association of Universities for Research in Astronomy, 
Inc., for NASA, under contract NAS 5-26555. 
\end{acknowledgements}

\end{document}